\documentclass[%
twocolumn,
amsmath,amssymb,
aps,
prx,
]{revtex4-2}

\usepackage{url}
\usepackage{amssymb}
\usepackage{amsfonts}
\usepackage{graphicx,epsfig,epstopdf}
\usepackage{soul}
\usepackage{amsmath}
\usepackage{color}
\usepackage[table]{xcolor}
\usepackage{multirow}
\usepackage{booktabs}
\usepackage[utf8]{inputenc}
\usepackage{upgreek}
\usepackage{caption}
\usepackage{subfigure}
\usepackage{hyperref}
\usepackage{array}
\usepackage{tabularx}
\usepackage{multirow}
\usepackage{pifont}
\newcommand{\cmark}{\ding{51}}%
\newcommand{\xmark}{\ding{55}}%
\usepackage{gensymb}
\newcolumntype{L}[1]{>{\raggedright\arraybackslash}p{#1}}
\newcolumntype{C}[1]{>{\centering\arraybackslash}p{#1}}
\newcolumntype{M}[1]{>{\centering\arraybackslash}m{#1}}

\usepackage{caption}
\usepackage{xcolor}
\usepackage{xpatch}

\makeatletter
\ExplSyntaxOn
\cs_new:Npn \bibColoredItems #1#2
  {
    \clist_map_inline:nn {#2} { \cs_new:cpn {bib@colored@##1} {#1} } 
  }
\ExplSyntaxOff

\newcommand\bib@setcolor[1]{%
  \ifcsname bib@colored@#1\endcsname
    \expandafter\color\expandafter{\csname bib@colored@#1\endcsname}
  \else
    \normalcolor
  \fi
}

\xpatchcmd\@bibitem
  {\item}
  {\bib@setcolor{#1}\item}
  {}{\fail}

\xpatchcmd\@lbibitem
  {\item}
  {\bib@setcolor{#2}\item}
  {}{\fail}

\def\e{\begin{equation}}
\def\f{\end{equation}}
\def\=#1{\overline{\overline #1}}
\def\*#1{\overline{\overline{\overline #1}}}

\def\_#1{{\bf #1}}

\def\.{\cdot}

\def\##1{{\bf#1\mit}}
\def\Re{{\rm Re\mit}}
\def\Im{{\rm Im\mit}}
\def\l#1{\label{eq:#1}}
\def\r#1{(\ref{eq:#1})}

\begin{document}




\title{Classification of bianisotropic metasurfaces\\ from reflectance and transmittance measurements}
 
\author{
M.~Albooyeh$^{1,2,\ast}$, V. Asadchy$^{3,4}$, J. Zeng$^{1,5}$, M. Rajaee$^1$, H. Kazemi$^1$, 
M. Hanifeh$^1$, and F. Capolino$^1$}
 
\affiliation{$^1$Department of Electrical Engineering and Computer Science, University of California, Irvine, CA 92697, USA}
\affiliation{$^2$Mobix Labs Inc., 15420 Laguna Canyon, Irvine, California 92618, USA}
\affiliation{$^3$Department  of  Electrical  Engineering,  Stanford  University,  Stanford, CA  94305, USA}
\affiliation{$^4$Department of Electronics and Nanoengineering, Aalto University, P.O.~Box 15500, FI-00076 Aalto, Finland}
\affiliation{$^5$Wuhan National Laboratory for Optoelectronics, Huazhong University of Science and Technology, Wuhan, 430074, Hubei, China}

\address{\rm{$^\ast$corresponding author: mohammad.albooyeh@gmail.com}}

\begin{abstract}
Upon using fundamental electromagnetic properties of metasurfaces we build a platform  to classify reciprocal bianisotropic metasurfaces from typical experimental measurements and determine isotropic, anisotropic, bi-isotropic (chiral), and bianisotropic (so-called omega) properties. We provide experimental guidelines to identify each class by measuring macroscopic scattering parameters, i.e., reflection and transmission coefficients upon plane wave illumination  with linear and/or circular polarization. We explicitly provide a recipe of what metasurface properties can and cannot be inferred by means of chosen polarization, reflection, and transmission properties. We also clarify  common confusions in the classification of anisotropic versus chiral metasurfaces based on circular dichroism measurements presented in the recent literature.

\end{abstract}

\maketitle



\section{Introduction}\label{chiral2d}

An object is called chiral when it cannot be superimposed on its mirror image or, in other words, when its symmetry group does not possess any of the following elements: centre of inversion, reflection planes or rotation–reflection axes (see Refs.~\cite{kelvin619} and~\cite[Sec. 1.9.1]{Barron_mol}). This is a geometric definition of chirality. When interacting with  electromagnetic (EM) waves, geometrically chiral structures manifest themselves through the effects of  optical activity (OA) and/or circular dichroism (CD) resulting from circular birefringence~\cite{Barron_mol}. 
Following a famous discovery by Pasteur~\cite{Pasteur}, it was believed  for long time that   circular birefringence (reciprocal OA and CD) occurs only in materials whose microscopic constituents   are   chiral. However, that statement is only true for materials consisting of isotropic ensembles of constituents, as it was demonstrated in early studies during the mid-twentieth century~\cite{Bunn,Hobden,Williams0,Williams} and further developed later on in  1990s~\cite{saadoun,arnaut,sochava}. {OA and CD can occur even in materials with   achiral constituents due to the lattice asymmetry~\cite{kruk2015polarization} or due to  specific symmetry of the constituents (lacking a centre of inversion but possessing reflection planes or a rotation–reflection axis, see Ref.~\cite[Sec.~1.9.1]{Barron_mol}). 
The latter scenario  is  referred to as pseudochirality (or extrinsic chirality) and it will be discussed in more detail in Section~\ref{conorg} for metasurfaces.}  There is a large number of studies devoted to pseudochiral metasurfaces (e.g., \cite{saadoun,plum2009,Viki_mod}). Circular birefringence  induced by such metasurfaces occurs only for \textit{specific} angles of incidence.
Naturally, if the inclusions in such materials are randomly oriented, they do not exhibit electromagnetic  chirality (EC) for any incident light.
%
Thus, it is important to determine a characteristic signature  of  EC and how it can be estimated in an unknown material structure from optical measurements.

EC finds a variety of applications in stereochemistry, spectroscopy, drug engineering, optical components and displays, and nonlinear optics. Recently, there has been growing interest on chiral two-dimensional systems such as  planar arrays of scatterers and metasurfaces. In contrast to bulk materials, such as  solutions of chiral molecules or some non-centrosymmetric crystals,   planar arrays of scatterers and metasurfaces   can exhibit strong  EC even with sub-wavelength thicknesses. Typical unit cells of these structures are metallic or dielectric helices~\cite{Wegener,Rajaei1}, artificial liquid crystals~\cite{Zhao}, tetramers~\cite{Govorov}, etc. 
Noticeable EC can be observed even from scattering by centrosymmetric meta-atoms when they are sandwiched between two materials (substrate and superstrate) with high dielectric contrast. Such an effect is referred to as substrate-induced chirality~\cite{papakostas2003optical,kuwata,Banzer}. In this case, the unit-cell cannot be superimposed on its mirror image due to  the material contrast asymmetry. A flat spiral on a dielectric substrate is an example of this case.

The non-obvious relation between EC  and circular birefringence may yield serious misconceptions in the design of planar metasurfaces and arrays of scatterers and in the interpretation of their scattering properties. Recently  there was a large number of studies (e.g.,~\cite{weimin, arteaga2016relation, zhukovsky2009elliptical, valev2009plasmonic, novitsky2012asymmetric})  reporting EC for normally incident light on certain planar structures that, in fact, are not   chiral. Indeed, as will be shown below, reported CD in these works corresponds instead to anisotropy. Generally, CD can be generated in two ways: due to chirality and/or due to anisotropy. Moreover, in anisotropic materials, linear birefringence~\cite{Kaminsky} can be mistaken for EC.

Similarly to metasurfaces with chiral inclusions that exhibit unequal responses for circularly polarized waves with opposite handedness and are referred to as chiral metasurfaces, there is another class of metasurfaces that provides different responses when  illuminating waves are incident from opposite sides on the metasurface. Metasurfaces belonging to this class are composed of omega inclusions, from the shape of  the Greek letter $\Omega$, and are referred to as ``omega'' metasurfaces~\cite{biama,Albooyeh,Yazdi2,VAsadchy}.  In electromagnetism, any dipolar particle that simultaneously exhibits a magnetic dipole moment generated by the illuminating \textit{orthogonal} electric field and an electric dipole moment generated by the illuminating \textit{orthogonal} magnetic field is referred to as  omega-type since historically the noted condition holds for a wired omega-shape inclusion~\cite{Simovski}. We define an object to have  geometric ``omega'' symmetry if  the normal vectors to its mirror-symmetry planes are all parallel to a single plane~\cite{omega_def}. Here, a reflection (or a mirror-symmetry) plane is a plane that divides an object into halves that are related to each other as an object is to its mirror image~\cite[Sec.~2.2.4]{Ladd}. The lack of a rigorous analysis of the responses from chiral and omega canonical metasurfaces, as well as other {\it reciprocal} bianisotropic metasurfaces, to EM waves with arbitrary polarization, restricts the ability to investigate the range of applications and potentials that these metasurfaces provide (see footnotes~\cite{exp_bia, non_rec} for definitions of bianisotropy and reciprocity). Such rigorous analysis is the motivation of this work.

In this paper, we present a comprehensive study that clarifies the use of different terminologies in metasurface classification. It should be noted that such classification and analysis to some extent  was presented in Refs.~\cite{biama,Kaminsky} for bulk materials, and in {\color{black}{Refs.~\cite{pfeiffer2014bianisotropic, VAsadchy, Albooyeh,achouri2019angular,achouri2021fundamental,menzel2010advanced} }} for metasurfaces. Nevertheless, there is no rigorous study to date on how to determine anisotropic, chiral, or omega properties of metasurfaces from typical experimental measurements. In this study, we rigorously analyze  reciprocal metasurfaces composed of unit cells modeled with effective polarizabilities. We emphasize that any EM property of the inclusions (e.g. chirality, anisotropy, etc.) is extended to a similar EM property of a metasurface composed of a periodic or amorphous arrangement of such inclusions whereas the reverse is not necessarily true. For instance, a metasurface with rectangular unit-cells and composed of isotropic inclusions is anisotropic (assuming finite periodicity)~\cite{anis_def}.

As a final note, we emphasize that the word ``chirality'' is sometimes attributed to a property of EM field itself, i.e., fields can also be ``chiral'', i.e., they possess helicity (see e.g., Refs.~\cite[p.~41]{Barron} and~\cite{Tang1,Rajaei1,Mina1,Mina2}). To clarify, in this study we consider solely material chirality and analyze its interaction with EM waves. We note that in some studies (see e.g. Refs.~\cite{eftekhari, liu2014spontaneous, schwanecke, drezet, singh, xu, pfeiffer2014bianisotropic, xiong, huang, shi, schaferling}), the word chirality refers to the reflected and/or transmitted EM wave from {\it achiral} planar structures and is not referred to material. In the next section and before presenting an analytical model for the analysis of metasurfaces, we discuss a possible fallacy that may occur in the determination of possible EC of a planar object, which was the initial motivation for the present study.

\begin{figure*}
\centering
\begin{tabular}{cccc}
  \subfigure[]{
  \epsfig{file=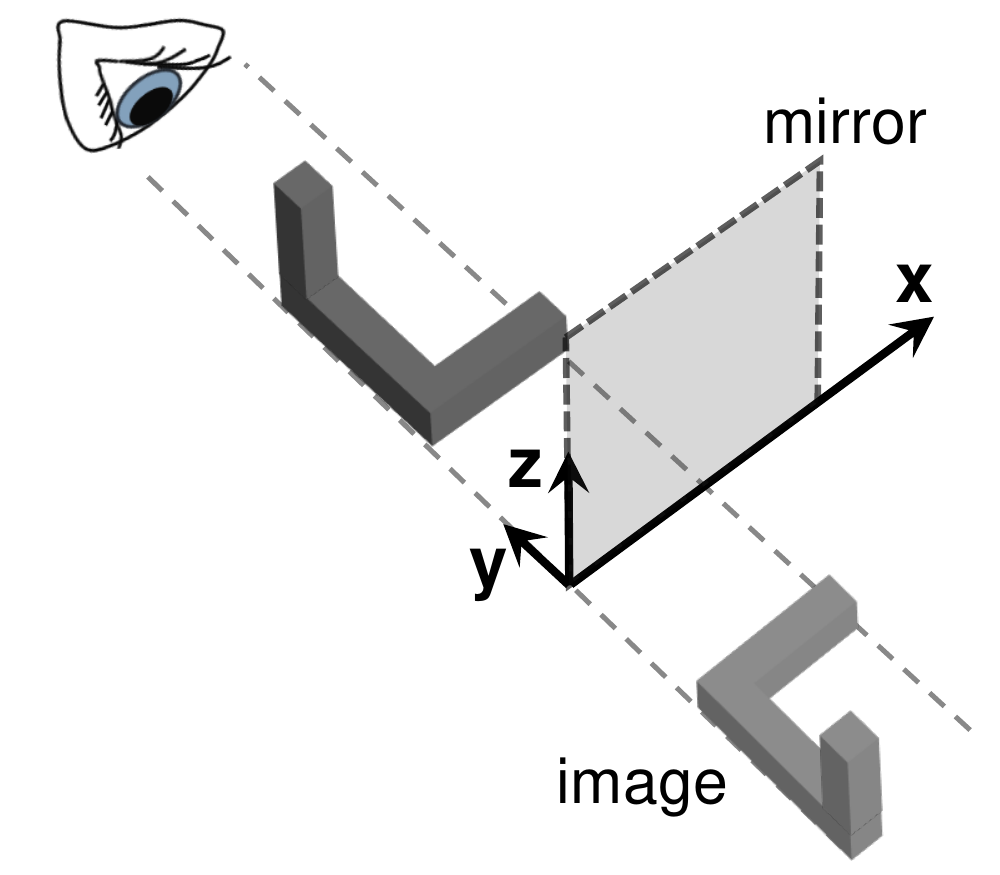, width=0.4\columnwidth}
  \label{ris:fig0e} }&
     \subfigure[]{
  \epsfig{file=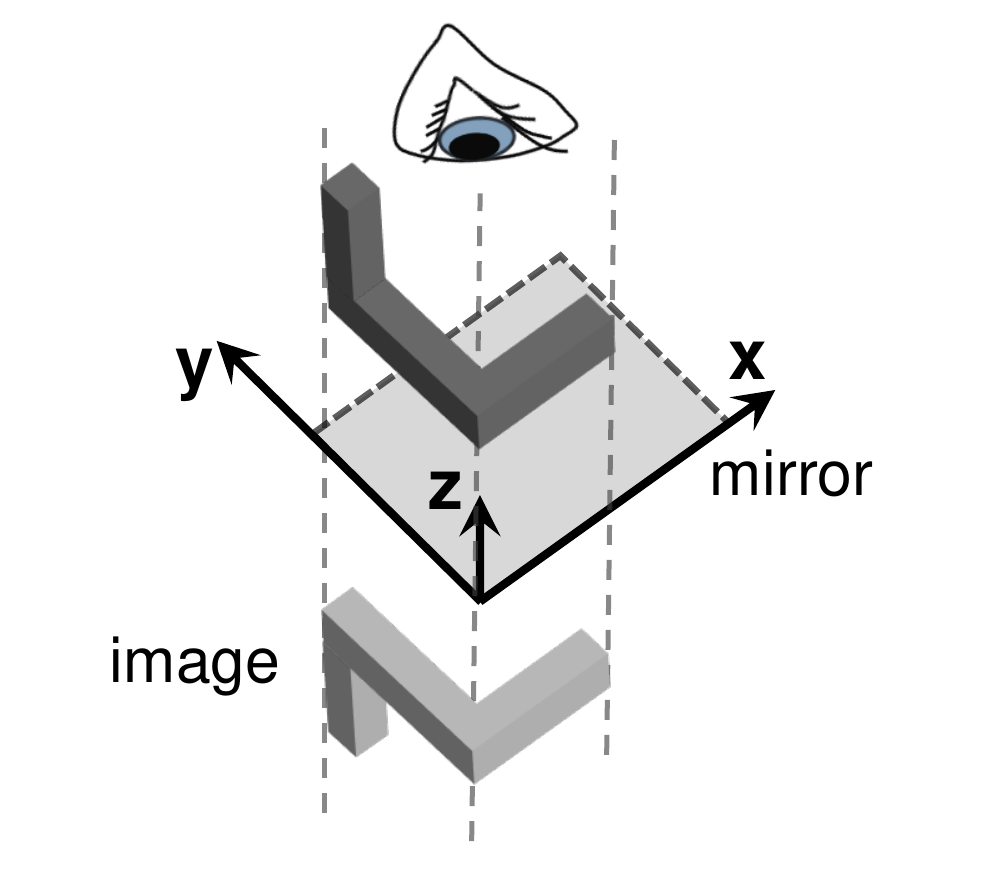, width=0.4\columnwidth}
  \label{ris:fig0f} }&
  \subfigure[]{
  \epsfig{file=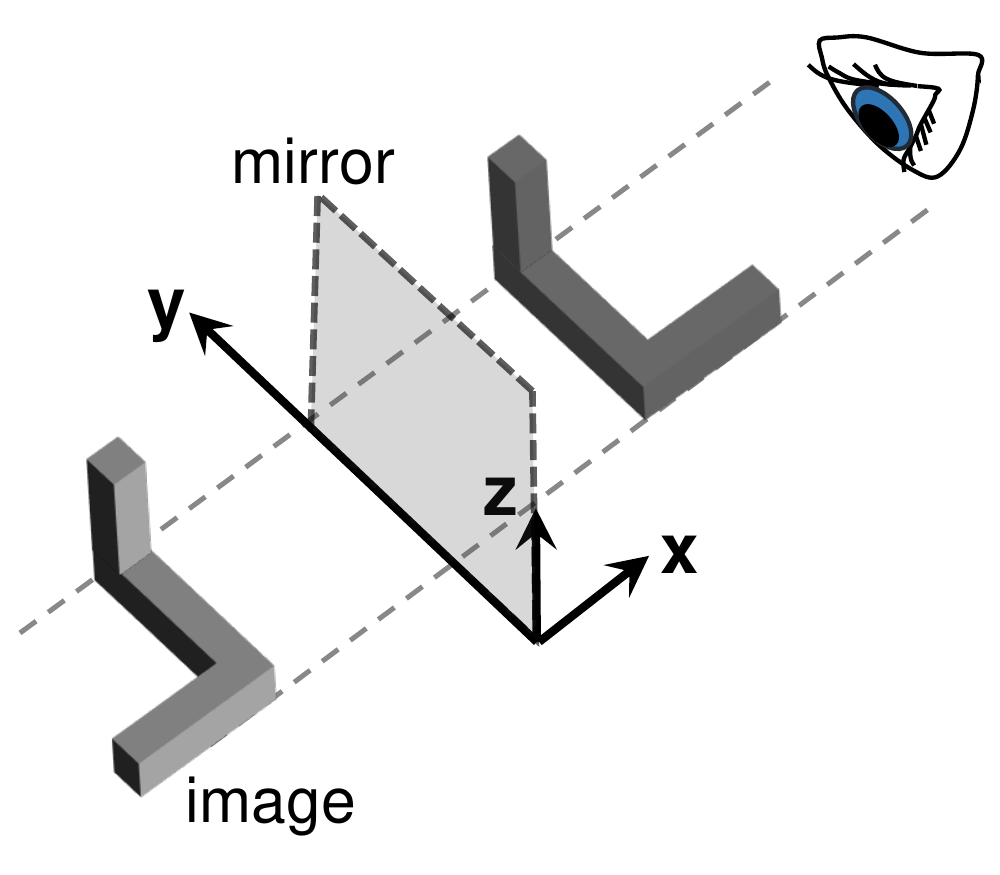, width=0.4\columnwidth}
  \label{ris:fig0g} }
  &
  \subfigure[]{
  \epsfig{file=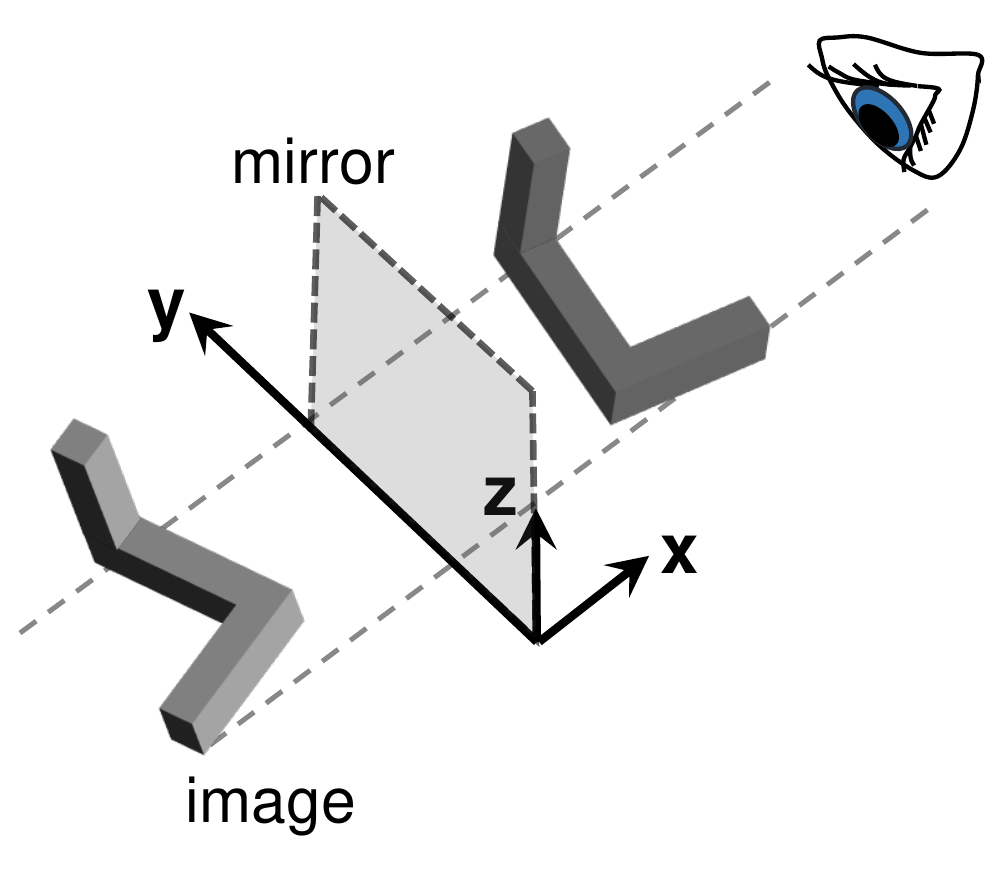, width=0.4\columnwidth}
  \label{ris:fig0h}}\\
  \subfigure[]{
  \epsfig{file=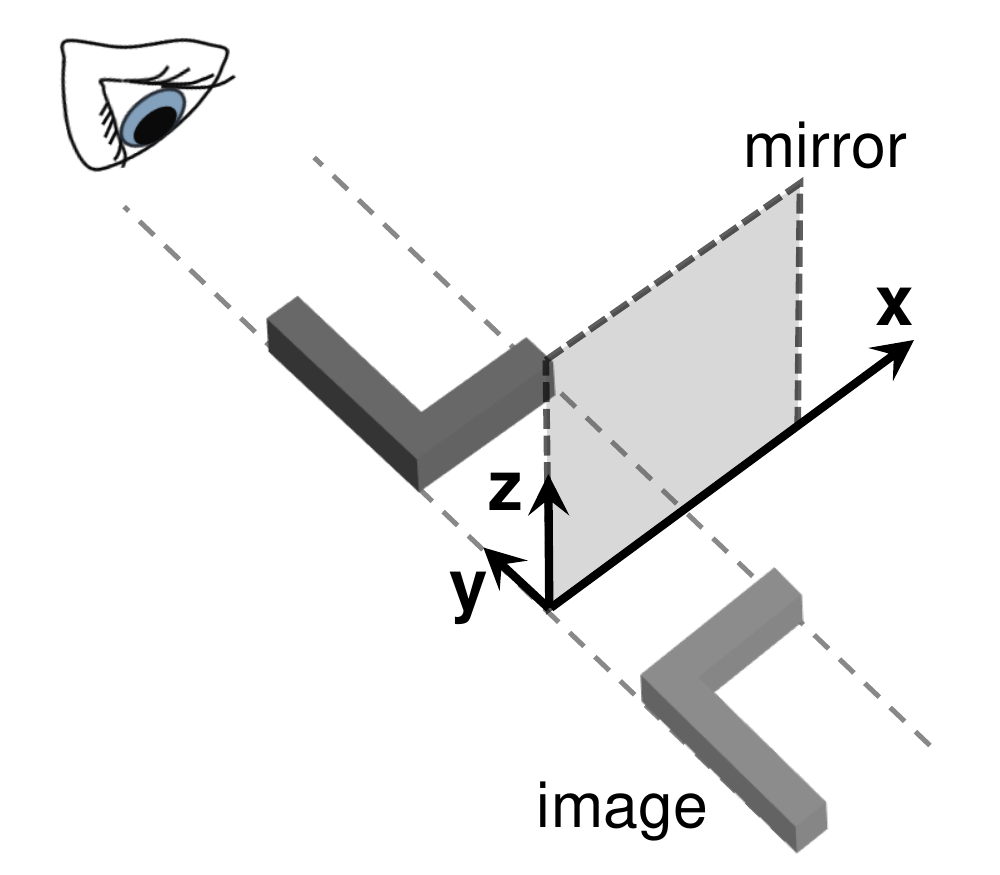, width=0.4\columnwidth}
  \label{ris:fig0a} }&
  \subfigure[]{
  \epsfig{file=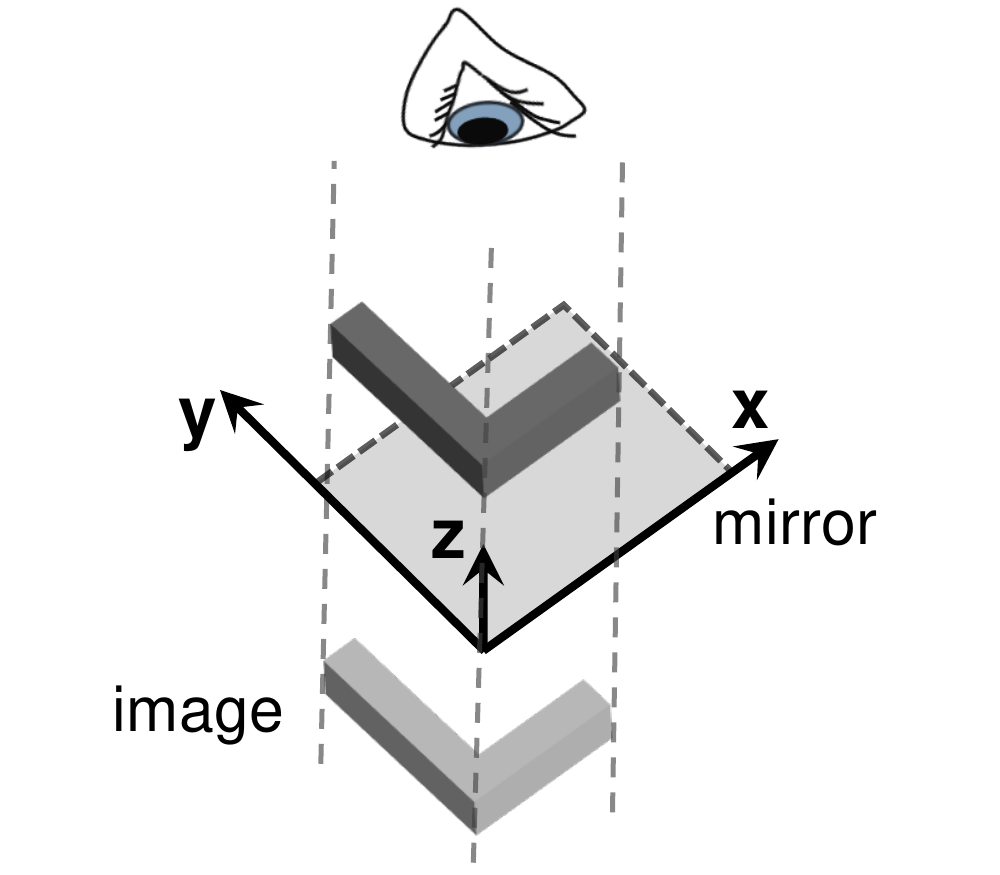, width=0.4\columnwidth}
  \label{ris:fig0b} }&
    \subfigure[]{
  \epsfig{file=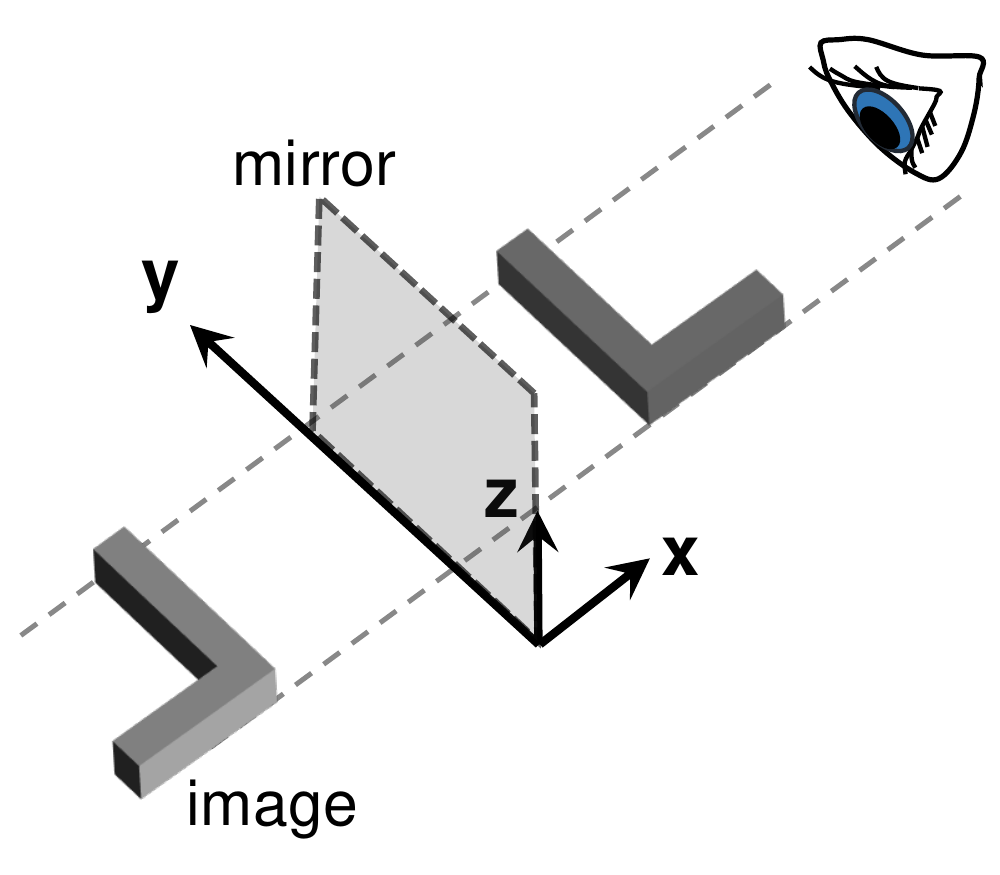, width=0.4\columnwidth}
  \label{ris:fig0c} }&
    \subfigure[]{
  \epsfig{file=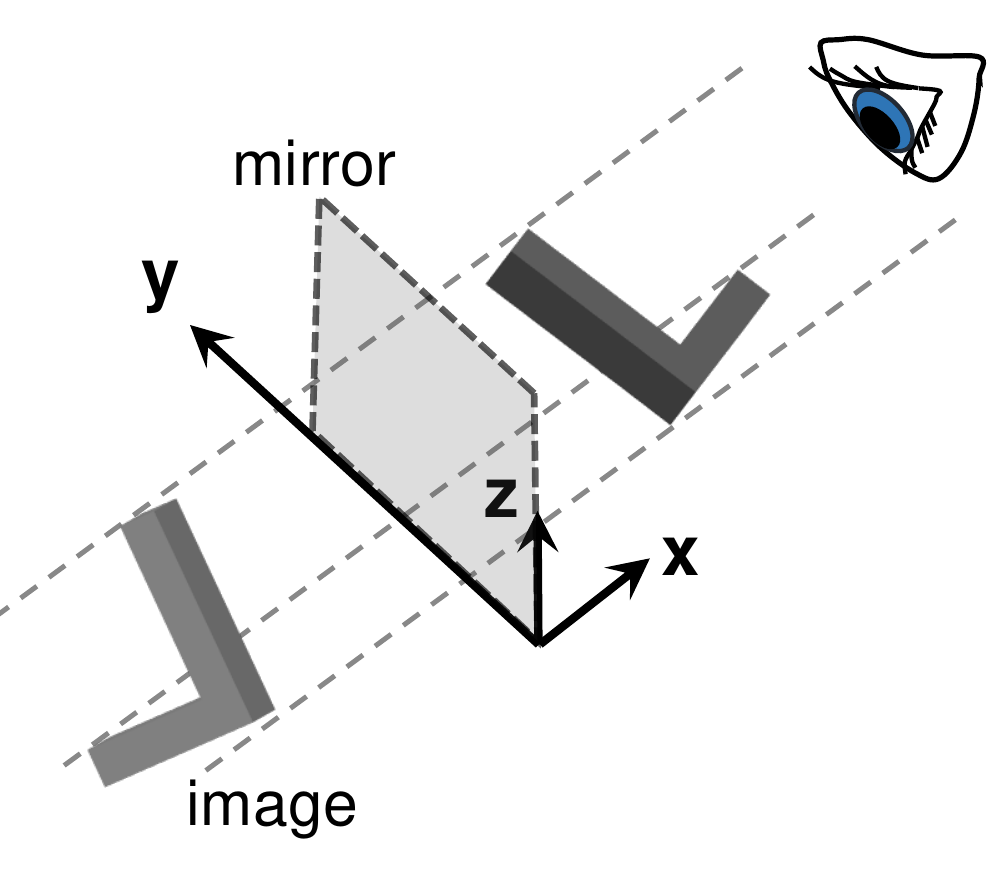, width=0.35\columnwidth}
  \label{ris:fig0d} }
\end{tabular} 
  \caption{(a)--(d) Twisted U-shaped particle in front of a mirror. Its image is not supperimposable on its original particle with a simple rotation and/or translation operation. Its EM response corresponds to that of a chiral particle for any illumination directions. The particle spatially expands in three orthogonal directions, i.e., it is a three-dimensional particle. (e)--(h)  L-shaped particle positioned in front of a mirror. In (e) and (g) the particle can be superimposed on its mirror-image by a simple $\pm180^\circ$ rotation around the $x$ and $y$ axes, respectively, whereas in (f) it can be superimposed on its mirror image by a simple translation along the $z$ axis.  This L-shaped particle is referred to as a {\it pseudochiral} particle and it exhibits  chiral properties only for   specific illumination directions such as in (h). }\label{ris:fig0} 
\end{figure*}

\section{Chirality and pseudochirality}\label{conorg}

Let us first consider a twisted U-shaped particle as an illustrative example of a geometrically chiral object, together with its mirror-image. Figures~\ref{ris:fig0e}--\ref{ris:fig0h}  
clearly show that the mirror-image  cannot  be superimposed (by rotation and/or translation) on the original object regardless of the position of the mirror. A material/metasurface composed of such geometrically chiral particles exhibits  EC for any illumination direction (note that if the material  is composed of a periodic or an amorphous array of such chiral inclusions, EC can be substantially minimized for illuminations at some angles~\cite[Suppl. Mat.]{Viki_mod}). Naturally,  a geometrically chiral particle cannot be two-dimensional (i.e., flat). 

Let us next consider the planar L-shaped particle shown in Figs.~\ref{ris:fig0a}--\ref{ris:fig0d}.  Note that in this discussion we assume a vanishing thickness of the particle (e.g. the L-shaped particle in Fig.~\ref{ris:fig0a} has a vanishing thickness along the $z$-direction, assumption valid when the thickness is very small compared to both the wavelength and the other two dimensions in the $x$ and $y$-directions). This L-shaped object has one mirror-symmetry plane, namely the $xy$-plane. Such L-shaped particles, as well as any other planar particles, are always superimposable on their mirror image by using simple rotation and/or translation operations (see Fig.~\ref{ris:fig0} and its caption). 

Note that whereas the  L-shaped particle in Fig.~\ref{ris:fig0b} is superimposable on its mirror-image by a simple translation along the $z$-axis, those in Figs.~\ref{ris:fig0a} and~\ref{ris:fig0c} are only superimposable on their mirror image by a $\pm 180^\circ$ rotation about the $x$- and $y$-axis, respectively.  

The L-shaped particle is called "pseudochiral" since it can exhibit similar effects to an EM chiral material response only for specific illumination directions~\cite{biama,plum2009,saadoun,sochava,Viki_mod}. Indeed, the simple L-shaped particle appears as a part of a helix to  light beams propagating along directions \textit{not contained} in (or \textit{perpendicular} to) its reflection-plane~\cite[Sec.~1.9.1]{Barron_mol}, as illustrated in Fig.~\ref{ris:fig0d}. This statement is in fact valid for  arbitrary pseudochiral dipolar particles, as we prove it rigorously using symmetry principles in  Supplementary Material~S--1.
Thus, the L-shaped particle may falsely be understood as a   chiral object if one restricts illumination and observation to certain specific directions only. A regular metasurface (or a planar array) composed of these L-shaped particles   may or may not generate OA and/or CD effects, depending on the direction of  illumination.

In the rest of the paper, we focus on reciprocal arrays with subwavelength period (metasurfaces), that are optically thin and made of electric and magnetic dipolar inclusions which are of significant interest in modern optics. In general, constituents of a reciprocal metasurface can be bianisotropic~\cite{exp_bia}, including canonical types of chiral and omega inclusions.

\section{Analytical Modeling of bianisotropic metasurfaces}\label{ana_st}
{\color{black}{In this section and only for self-consistency reasons, we provide some basic information required to present our analytical approach. Similar formalism maybe found elsewhere in the literature (see e.g., in Refs.~\cite{pfeiffer2014bianisotropic, VAsadchy, Albooyeh,achouri2019angular,achouri2021fundamental,menzel2010advanced}).}} From an EM point of view, planar arrays and metasurfaces are conventionally characterized by  surface susceptibilities, collective polarizabilities, or surface impedances~\cite{holloway,Albooyeh,smith,scher,VAsdchy_bc}. By using collective polarizability tensors, in this section we briefly classify general metasurfaces composed of {\it reciprocal} bianisotropic dipole scatterers~\cite{biama,Viki_Tutorial}. {\color{black}{We emphasize that such classification to some extent was presented in Refs.~\cite{pfeiffer2014bianisotropic, VAsadchy, Albooyeh,achouri2019angular,achouri2021fundamental,menzel2010advanced} for metasurfaces.}} 

\begin{figure}
\centering
   \subfigure[]{
  \epsfig{file=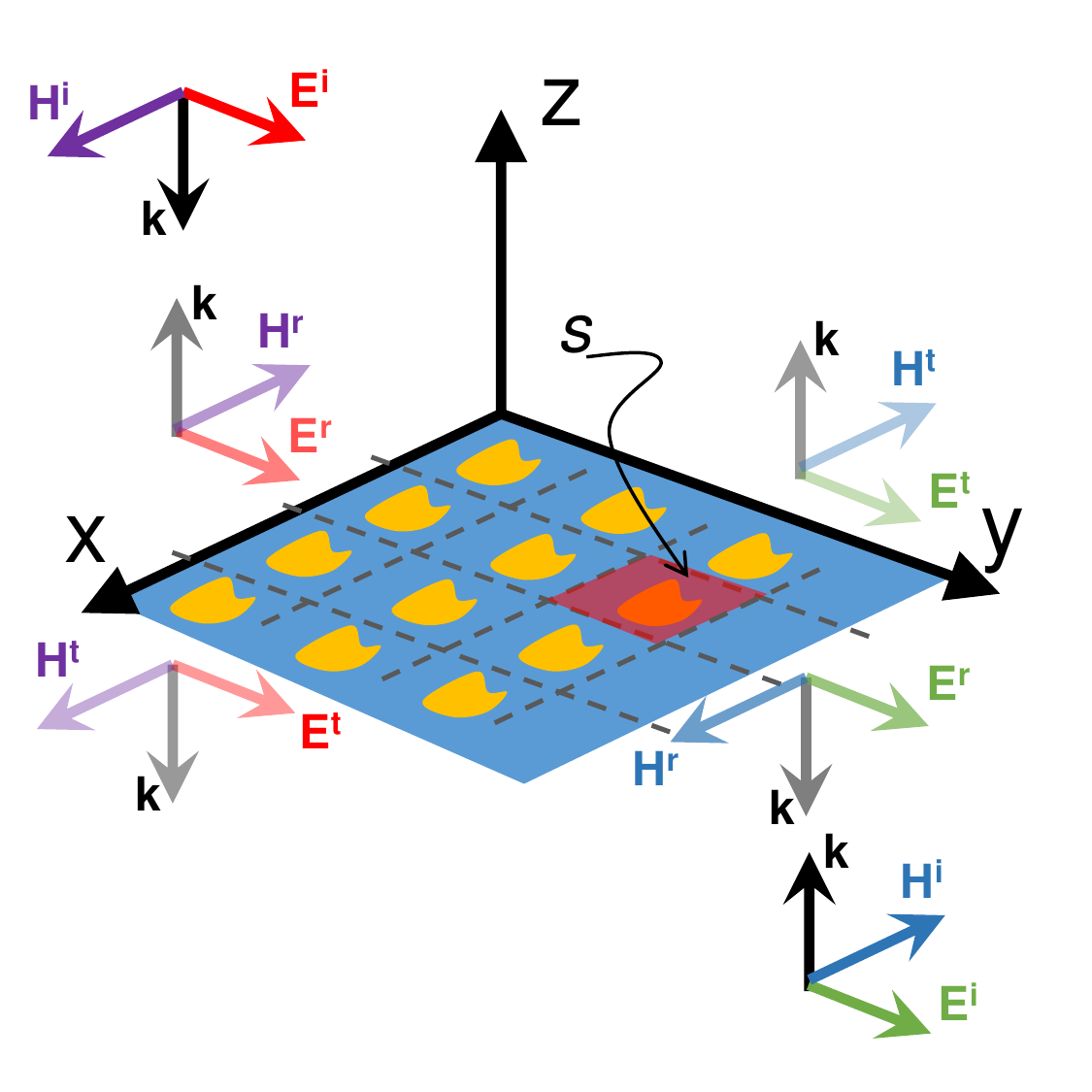, width=0.46\columnwidth}
   \label{ris:fig1a} }
     \subfigure[]{
   \epsfig{file=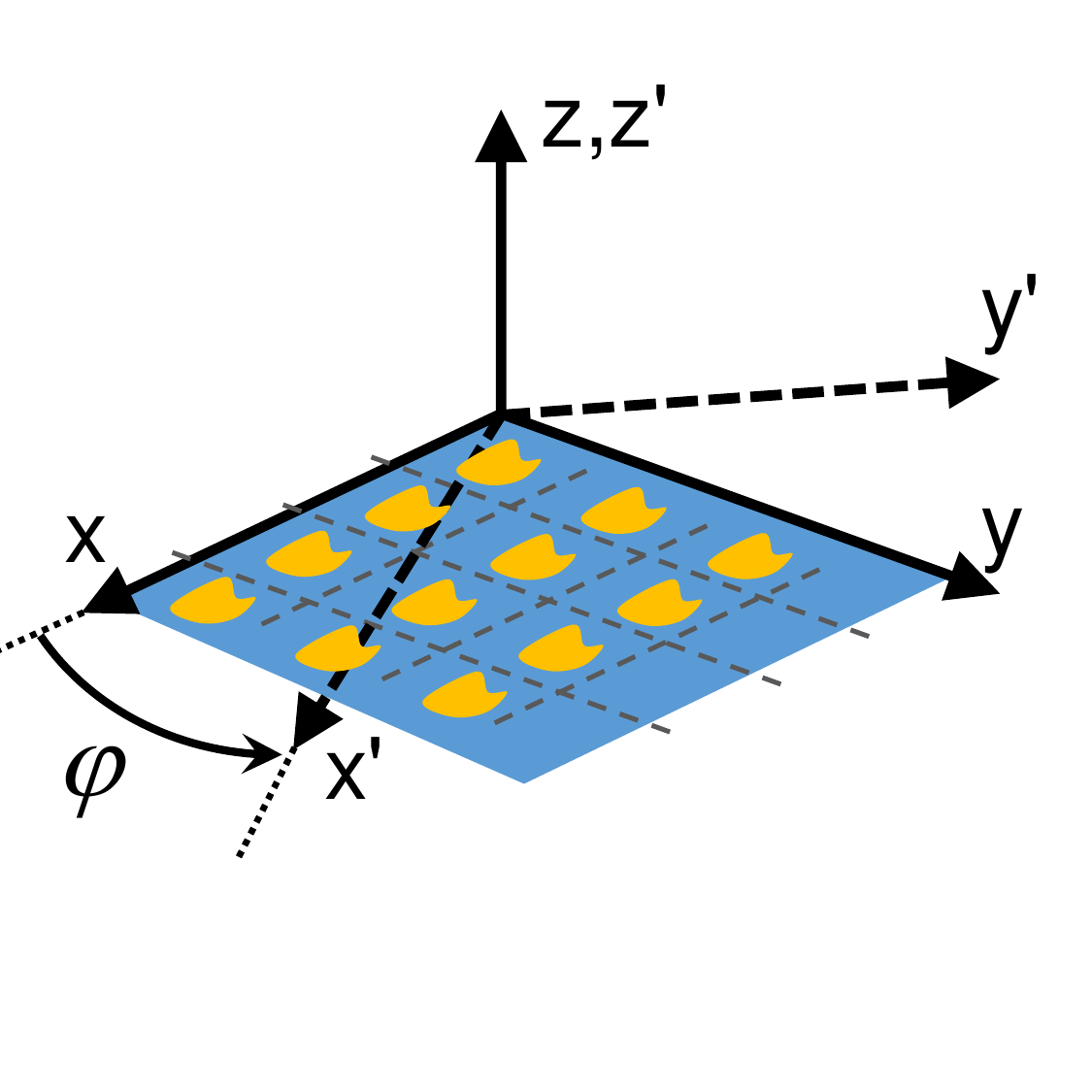, width=0.46\columnwidth}
   \label{ris:fig1b} }\\ 
  \caption{Periodic arrangement of dipole scatterers forming a metasurface in the $xy$-plane. (a) Schematic with possible illumination from the top or bottom of the metasurface plane. (b) Schematic with rotation of the coordinate system around the $z$ axis, normal to the metasurface plane.}\label{chirality2D_Rev1:fig1}
\end{figure}

Without loss of generality, we consider an array or a metasurface located in the $xy$-plane [see Fig.~\ref{ris:fig1a}]. In the following we will refer to them only as metasurfaces, assuming that with due caution the formalism in this paper can also be applied to arrays. 
 Here, we clarify what chirality, anisotropy, and omega properties imply in terms of  collective polarizability tensors, that is, in terms of metasurface characteristic parameters. The {\it constitutive} relations between the incident electric $ {\bf E}^\textrm{i}$ and magnetic $ {\bf H}^\textrm{i}$ fields  and the induced electric $ {\bf P}$ and magnetic $ {\bf M}$ equivalent surface polarization densities of an arbitrary bianisotropic metasurface are~\cite{Niemi, Albooyeh}
\begin{eqnarray}
  \nonumber {{\bf   P}} &=& {\bar{\bar{\alpha}}^{\textrm{ee}}\over S} \cdot {\bf E}^\textrm{i}+{\bar{\bar{\alpha}}^{\textrm{em}}\over S} \cdot {\bf H}^\textrm{i}, \\
  {{\bf   M}} &=& {\bar{\bar{\alpha}}^{\textrm{me}}\over S} \cdot {\bf E}^\textrm{i}+{\bar{\bar{\alpha}}^{\textrm{mm}} \over S} \cdot {\bf H}^\textrm{i}, \label{chirality2D_Rev1:our_collective}
\end{eqnarray}
respectively, where $S$ is the unit cell area [see Fig.~\ref{ris:fig1a}]. Furthermore, $\bar{\bar{\alpha}}^{\textrm{ee}}$, $\bar{\bar{\alpha}}^{\textrm{mm}}$, $\bar{\bar{\alpha}}^{\textrm{em}}$, and $\bar{\bar{\alpha}}^{\textrm{me}}$  are, respectively, the electric, magnetic, magnetoelectric, and electromagnetic \textit{collective polarizability} tensors. The latter two terms, $\bar{\bar{\alpha}}^{\textrm{em}}$ and $\bar{\bar{\alpha}}^{\textrm{me}}$, are representing the collective  bianisotropic properties~\cite{biama,exp_bia} associated to the  metasurface unit cell. We call them ``collective'' since they account for the polarizability of each individual array inclusion as well as the interactions among all the inclusions. The reader is referred to Sec.~S--2 in Supplementary Material for a discussion about the units of  polarizabilities.

{\color{black} The   homogeneization theory describing the metasurface based on  collective  dipolar polarizabilities is valid also for oblique angle of incidence, since this information is included in the coupling effects. This means that the polarizabilities in Eqs.~(\ref{chirality2D_Rev1:our_collective}) would account for such direction of incidence. Indeed, an analogous homogenization method for bulk materials made of a lattice of dipoles including spatial dispersion provided in Refs.~\cite{silveirinha2007generalized,Lannebere:14}, which can be straightforwardly applied to the case of the homogeneization of a metasurface with oblique incidence. Nevertheless, in this paper we focus to normal incidence since it is the most important case for the proposed characterization. A discussion about the validity to model the metasurface elements using dipoles can be found in Ref.~\cite{kruk2015polarization}.}

Note that each polarizability tensor in Eqs.~(\ref{chirality2D_Rev1:our_collective}) generally consists of nine components. However, the illumination type determines which of the properties are contributing to the reflected and transmitted fields. For example, when using a plane wave normal illumination, the normal polarizability components (to the metasurface plane) are not contributing to the reflected/transmitted waves. Therefore, to simplify our discussion on the metasurface EM  properties we specify here our excitation types. We assume that the metasurface is illuminated by a plane wave [with either linear polarization (LP) or circular polarization (CP)]. The incident wave is either impinging on the metasurface from the bottom or top side [see Fig.~\ref{ris:fig1a}] and we only study normal (to the metasurface plane) incidence. 

As a result of our illumination selection, i.e., normal illumination, for each polarizability tensor in Eqs.~(\ref{chirality2D_Rev1:our_collective}), only four transversal components out of the nine components contribute to the reflected/transmitted fields and we only consider these components in our classification and analysis. For instance, in Cartesian coordinates we consider each polarizability tensor with the following components $\bar{\bar{{\alpha}}}=\alpha_{xx} \hat{{\bf x}} \hat{{\bf x}}+\alpha_{xy} \hat{{\bf x}}\hat{{\bf y}}+\alpha_{yx} \hat{{\bf y}}\hat{{\bf x}}+\alpha_{yy} \hat{{\bf y}}\hat{{\bf y}}$,
where $\hat{{\bf x}}$ and $\hat{{\bf y}}$ are the unit vectors in the Cartesian coordinate system, moreover, $\hat{{\bf x}}\hat{{\bf x}}$, $\hat{{\bf y}}\hat{{\bf y}}$, $\hat{{\bf y}}\hat{{\bf x}}$, and $\hat{{\bf x}}\hat{{\bf y}}$ are the dyadic products of unit vectors.  When using Cartesian coordinates, the tensor is equivalent to a $2\times 2$ matrix $\bar{\bar{{\alpha}}}=\left[ \alpha_{xx}, \alpha_{xy} ; \alpha_{yx}, \alpha_{yy}  \right]$.
 For reciprocal metasurfaces (as considered throughout this work), the tensors satisfy the relations $\bar{\bar{{\alpha}}}^{\textrm{ee}}=\left({\bar{\bar{\alpha}}}^{\textrm{ee}}\right)^{T}$, $\bar{\bar{{\alpha}}}^{\textrm{mm}}=\left({\bar{\bar{\alpha}}}^{\textrm{mm}}\right)^{T}$, and $\bar{\bar{{\alpha}}}^{\textrm{em}}=-\left({\bar{\bar{\alpha}}}^{\textrm{me}}\right)^{T}$, where the superscript ``T'' stands for transpose operation [see Sec. 3.3.1 of Ref.~\onlinecite{biama} for a detailed discussion].
In summary, to fully characterize a reciprocal metasurface located in the $xy$-plane and illuminated at normal incidence, we require the following components of polarizability tensors: ${\alpha}^{\textrm{ee}}_{ii}$, ${\alpha}^{\textrm{ee}}_{jj}$, ${\alpha}^{\textrm{ee}}_{ij}$, ${\alpha}^{\textrm{mm}}_{ii}$, ${\alpha}^{\textrm{mm}}_{jj}$, ${\alpha}^{\textrm{mm}}_{ji}$, and all of four components of tensor $\bar{\bar{\alpha}}^{\textrm{em}}$, i.e., ${\alpha}^{\textrm{em}}_{ii}$, ${\alpha}^{\textrm{em}}_{ij}$, ${\alpha}^{\textrm{em}}_{ji}$, and ${\alpha}^{\textrm{em}}_{jj}$, where $i$ and $j$ are general indices in the $xy$-plane which may correspond to any prime coordinates rotated about the $z$-axis [see Fig.~\ref{chirality2D_Rev1:fig1}(b) and Sec.~S--6 in Supplementary Material for a detailed discussion]. We now have all the tools to classify general reciprocal bianisotropic metasurfaces in terms of anisotropic, chiral, and omega metasurface properties. Indeed, since we only consider the transverse polarizability components for a 2D surface, discussing isotropy in a general 3D sense would not be meaningful. As a result, we consider the following EM classes for reciprocal metasurfaces


\begin{table*}[t]
\centering
\caption{Reciprocal metasurface classes based on their (bi-)anisotropy properties in terms of in-plane polarizability components, and also reflection and transmission coefficients in both LP and CP bases.}
\label{Tab1}
\resizebox{\textwidth}{!}
{
\begin{tabular}{|c|c|c|c|c|c|c|c|c|}
\hline
\multirow{2}{*}{\bf class} & \multicolumn{2}{c|}{\multirow{2}{*}{\bf ISOTROPIC}}  & \multicolumn{2}{c|}{\multirow{2}{*}{\bf GENERAL ANISOTROPIC}}   & \multicolumn{2}{c|}{\bf BI--ISOTROPIC} & \multicolumn{2}{c|}{\bf GENERAL BI--ANISOTROPIC}\\ \cline{6-9} & \multicolumn{2}{c|}{} & \multicolumn{2}{c|}{}  & \multicolumn{2}{c|}{{\bf CHIRAL}} & {\bf CHIRAL} & {\bf OMEGA}  \\
\hline
\hline
\multirow{5}{*}{\bf polarizability} & \multicolumn{2}{c|}{}  &  \multicolumn{2}{c|}{}  & \multicolumn{2}{c|}{}  & \multirow{2}{*}{} & \multirow{2}{*}{} \\
& \multicolumn{2}{c|}{ $\alpha^{\rm ee}_{ii}=\alpha^{\rm ee}_{jj}~~~~~~~$ $\alpha^{\rm ee}_{ij}=\alpha^{\rm ee}_{ji}=0$ } & \multicolumn{2}{c|}{ $\alpha^{\rm ee}_{ii} \neq \alpha^{\rm ee}_{jj}~~~~~$  $\alpha^{\rm ee}_{ij}=\alpha^{\rm ee}_{ji}\neq 0$ }& \multicolumn{2}{c|}{ $\alpha^{\rm em}_{ii}=\alpha^{\rm em}_{jj} \neq 0~~$ $\alpha^{\rm em}_{ij}=\alpha^{\rm em}_{ji}=0$ } & \multirow{5}{*}{ $\alpha^{\rm em}_{ii}\neq\alpha^{\rm em}_{jj}$} & \multirow{2}{*}{ $\alpha^{\rm em}_{ij}\neq 0$} \\
& \multicolumn{2}{c|}{}  &  \multicolumn{2}{c|}{}  &  \multicolumn{2}{c|}{}  & \multirow{2}{*}{} & \multirow{2}{*}{} \\  
&\multicolumn{2}{c|}{ {$\alpha^{\rm mm}_{ii}=\alpha^{\rm mm}_{jj}~~$} {\color{black}$\alpha^{\rm mm}_{ij}=\alpha^{\rm mm}_{ji}=0$} } &\multicolumn{2}{l|}{$\alpha^{\rm mm}_{ii} \neq \alpha^{\rm mm}_{jj}~~~$ {\color{black}$\alpha^{\rm mm}_{ij}=\alpha^{\rm mm}_{ji}\neq 0$} } & \multicolumn{2}{c|}{ {$\alpha^{\rm mm}_{ii}=\alpha^{\rm mm}_{jj}~~$} {\color{black}$\alpha^{\rm mm}_{ij}=\alpha^{\rm mm}_{ji}=0$} }  &  &  \\

  & \multicolumn{2}{c|}{}  & \multicolumn{2}{c|}{}  &  \multicolumn{2}{c|}{}  & \multirow{2}{*}{} & \multirow{2}{*}{} \\

  & \multicolumn{2}{c|}{$\bar{\bar{\alpha}}^{\rm em}=0$}  & \multicolumn{2}{c|}{$\bar{\bar{\alpha}}^{\rm em}=0$}  &  \multicolumn{2}{c|}{ $\alpha^{\rm ee}_{ii}=\alpha^{\rm ee}_{jj}~~~~~~~$ {\color{black}$\alpha^{\rm ee}_{ij}=\alpha^{\rm ee}_{ji}=0$} }  & \multirow{2}{*}{} & \multirow{1}{*}{\color{black}$\alpha^{\rm em}_{ji}\neq 0$} \\  

  & \multicolumn{2}{c|}{} &  \multicolumn{2}{c|}{}  &  \multicolumn{2}{c|}{}  & \multirow{2}{*}{} & \multirow{2}{*}{} \\
\hline
\hline
\multirow{5}{*}{$r$ and $t$, {\bf LP basis}} & \multicolumn{2}{c|}{}  & \multicolumn{2}{c|}{}  &  \multicolumn{2}{c|}{}  & \multirow{2}{*}{} & \multirow{2}{*}{} \\
 
 & \multicolumn{2}{c|}{ $t^{\pm}_{ii}=t^{\pm}_{jj}~~~~~~~~$   {\color{black}$t^{\pm}_{ij}=t^{\pm}_{ji}=0$} } & \multicolumn{2}{c|}{ $t^{\pm}_{ii} \neq t^{\pm}_{jj}~~~~~~~~$  {\color{black}$t^{\pm}_{ij}=t^{\pm}_{ji}$} } &\multicolumn{2}{c|}{ {\color{black}$t^{\pm}_{ii}= t^{\pm}_{jj}~~~~~$ $t^{\pm}_{ij} = t^{\pm}_{ji}\neq 0$} } & {\color{black}$t^{\pm}_{ij}\neq t^{\pm}_{ji} $} & \multicolumn{1}{c|}{\color{black}$r^{+}_{ii}\neq r^{-}_{ii}$} \\
 
  & \multicolumn{2}{c|}{}  &  \multicolumn{2}{c|}{} & \multicolumn{2}{c|}{} & \multirow{2}{*}{} & \multirow{2}{*}{} \\
  
& \multicolumn{2}{c|}{$r^{\pm}_{ii}=r^{\pm}_{jj}~~~~~~~~$ {\color{black}$r^{\pm}_{ij}=r^{\pm}_{ji}=0$} }  &\multicolumn{2}{c|}{$r^{\pm}_{ii} \neq r^{\pm}_{jj}~~~~~~~~$  {\color{black}$r^{\pm}_{ij}=r^{\pm}_{ji}$} }& \multicolumn{2}{c|} {\color{black}$r^{\pm}_{ii}= r^{\pm}_{jj}~~~$  $r^{\pm}_{ij} = r^{\pm}_{ji}= 0$} & \multicolumn{1}{c|} {\color{black}$r^{+}_{ij/ji} \neq r^{-}_{ij/ji}$} &  \multicolumn{1}{c|} {\color{black}$r^{+}_{jj} \neq r^{-}_{jj}$} \\
& \multicolumn{2}{c|}{} &  \multicolumn{2}{c|}{} &  \multicolumn{2}{c|}{} & \multirow{2}{*}{} & \multirow{2}{*}{} \\
\hline
\hline
\multirow{5}{*}{$r$ and $t$, {\bf CP basis}} & \multicolumn{2}{c|}{} &  \multicolumn{2}{c|}{} &  \multicolumn{2}{c|}{} & \multirow{2}{*}{} & \multirow{2}{*}{} \\
& \multicolumn{2}{c|}{ $t^{\pm}_{\rm RR}=t^{\pm}_{\rm LL}~~~~~~$ {\color{black}$t^{\pm}_{\rm RL}=t^{\pm}_{\rm LR}=0$} } & \multicolumn{2}{c|}{$t^{\pm}_{\rm RR} = t^{\pm}_{\rm LL}~~~~~~~~$  {\color{black}$t^{\pm}_{\rm RL}\neq t^{\pm}_{\rm LR}$} }& \multicolumn{2}{c|}{\color{black}$t^{\pm}_{\rm RR} \neq t^{\pm}_{\rm LL}~~~$ $t^{\pm}_{\rm RL} = t^{\pm}_{\rm LR}=0$} & {\color{black}$t^{\pm}_{\rm RR}\neq t^{\pm}_{\rm LL}$} & \multicolumn{1}{c|}{\color{black}$r^{+}_{\rm RR,LL} \neq r^{-}_{\rm RR,LL}$} \\
&  \multicolumn{2}{c|}{} &  \multicolumn{2}{c|}{} & \multicolumn{2}{c|}{} & \multirow{2}{*}{} & \multirow{2}{*}{} \\
& \multicolumn{2}{c|}{$r^{\pm}_{\rm RR}=r^{\pm}_{\rm LL}=0~~~~~$ {\color{black}$r^{\pm}_{\rm RL}=r^{\pm}_{\rm LR} $} } &\multicolumn{2}{c|}{$r^{\pm}_{\rm RR} \neq r^{\pm}_{\rm LL}~~~~~$ {\color{black}$r^{\pm}_{\rm RL}=r^{\pm}_{\rm LR}$} }& \multicolumn{2}{c|} {\color{black}$r^{\pm}_{\rm RR}=r^{\pm}_{\rm LL}=0~~~$ $r^{\pm}_{\rm RL}=r^{\pm}_{\rm LR}$} & \multicolumn{1}{c|} {\color{black}$r^{\pm}_{\rm RR} \neq r^{\pm}_{\rm LL}$} & \multicolumn{1}{c|} {\color{black}$r^{\pm}_{\rm RL}\neq r^{\mp}_{\rm LR}$} \\
&  \multicolumn{2}{c|}{} &  \multicolumn{2}{c|}{} &  \multicolumn{2}{c|}{} & \multirow{2}{*}{} & \multirow{2}{*}{} \\
\hline
\hline
\end{tabular}
}
\end{table*}


\begin{enumerate}
\item A metasurface is said to be electromagnetically isotropic (in its plane) if the collective polarizabilities are such that $\alpha^{\rm ee}_{ii}=\alpha^{\rm ee}_{jj}$, $\alpha^{\rm mm}_{ii}=\alpha^{\rm mm}_{jj}$, and $\alpha^{\rm ee}_{ij}=\alpha^{\rm ee}_{ji}=0$, $\alpha^{\rm mm}_{ij}=\alpha^{\rm mm}_{ji}=0$ and also all bianisotropic terms $\bar{\bar{{\alpha}}}^{\textrm{em}}$ and ${\bar{\bar{\alpha}}}^{\textrm{me}}$ are vanishing.

\item A metasurface is anisotropic if $\alpha_{ii} \neq \alpha_{jj}$ and/or $\alpha_{ij}=\alpha_{ji}\neq 0$ for all electric and magnetic polarizabilities while all bianisotropic terms $\bar{\bar{{\alpha}}}^{\textrm{em}}$ and ${\bar{\bar{\alpha}}}^{\textrm{me}}$ are vanishing.

\item A metasurface is bi-isotropic if in addition to conditions in item {\bf 1}, $\alpha^{\rm em}_{ii}=\alpha^{\rm em}_{jj} \neq 0$ and $\alpha^{\rm em}_{ij}=\alpha^{\rm em}_{ji} = 0$.

\item A metasurface is bianisotropic if $\alpha^{\rm em}_{ii}\neq\alpha^{\rm em}_{jj}$ and/or either $\alpha^{\rm em}_{ij}$ or $\alpha^{\rm em}_{ji}$ is non-vanishing.\begin{enumerate}
\item A bianisotropic metasurface contains chiral properties if either  $\alpha_{ii}^{\rm em}$ or $\alpha_{jj}^{\rm em}$ is {\color{black}non-}vanishing.
\item A bianisotropic metasurface contains omega properties if either $\alpha_{ij}^{\rm em}$ or $\alpha_{ji}^{\rm em}$ is nonzero.
\end{enumerate}
\end{enumerate}

According to above classification, the EM chiral property lies within the category of bianisotropic or bi-isotropic metasurfaces whereas an EM omega property lies only within the category of  bianisotropic metasurfaces. Also note that if the metasurface is isotropic in the $xy$-plane, then, the characteristic parameters that fully characterize the metasurface reduce to ${\alpha}^{\textrm{ee}}_{ii}$, ${\alpha}^{\textrm{ee}}_{ij}$, ${\alpha}^{\textrm{mm}}_{ii}$, ${\alpha}^{\textrm{mm}}_{ij}$, ${\alpha}^{\textrm{em}}_{ii}$, and ${\alpha}^{\textrm{em}}_{ij}$, i.e., only $6$ parameters. Note that by isotropy in a surface we mean in-plane isotropy (or C4 symmetry), i.e., the absence of any in-plane preferred direction. The first two rows in Table~\ref{Tab1} summarize the above discussion and represent the relation between metasurface properties (classes) and the polarizability components.


{\color{black} We further classify metasurfaces based on reflection and transmission data. Classification of metasurfaces based on complex scattering parameters (reflection and transmission) and the Jones matrices has already been presented in the literature to some extent~\cite{achouri2021fundamental,menzel2010advanced}. 
However, such a classification   is not practical in the optical frequency range where phase measurements   are very complicated.
Therefore, in this work, we intend to classify bianisotropic metasurfaces   based on ``amplitude-only'' data of reflection and transmission coefficients either in an LP basis or in a CP basis and for ``one-side-only'' illumination scenarios. For brevity, we skip the derivation of equations and refer the readers to Sec.~S--3 of the Supplementary Material and we only present  a summary here.} {\color{black} By using the analytical approach as presented in Sec.~S--3 of Supplementary Material one can find the reflection and transmission coefficients in an  LP or a CP basis in terms of the polarizability components [see  Eqs.~(S.5)--(S.12) in an LP basis  and Eqs.~(S.15)--(S.22) in a CP basis in Sec.~S--3 of Supplementary Material]. }
{\color{black}  Therefore, based on the analysis presented Sec.~S--3 of Supplementary Material, in Table~\ref{Tab1} we add two last rows that summarize different classes of metasurfaces in terms of the reflection and transmission coefficients in LP basis [Eqs.~(S.5)--(S.12)] and in CP basis [Eqs.~(S.15)--(S.22)]. }

Note that if the amplitudes and phases of {\it all} co and cross polarized reflection/transmission coefficients are known, then, it is possible to transform all of them  from LP to CP basis and vice-versa (see Sec.~S--8 in Supplementary Material, and also Refs.~\cite{Corbaton,Cuesta}). However, in general both amplitude and phases may not be easily accessible, especially at optical frequencies, therefore, there is a fundamental difference working in LP or CP basis. 

\begin{figure}
\centering
{\epsfig{file=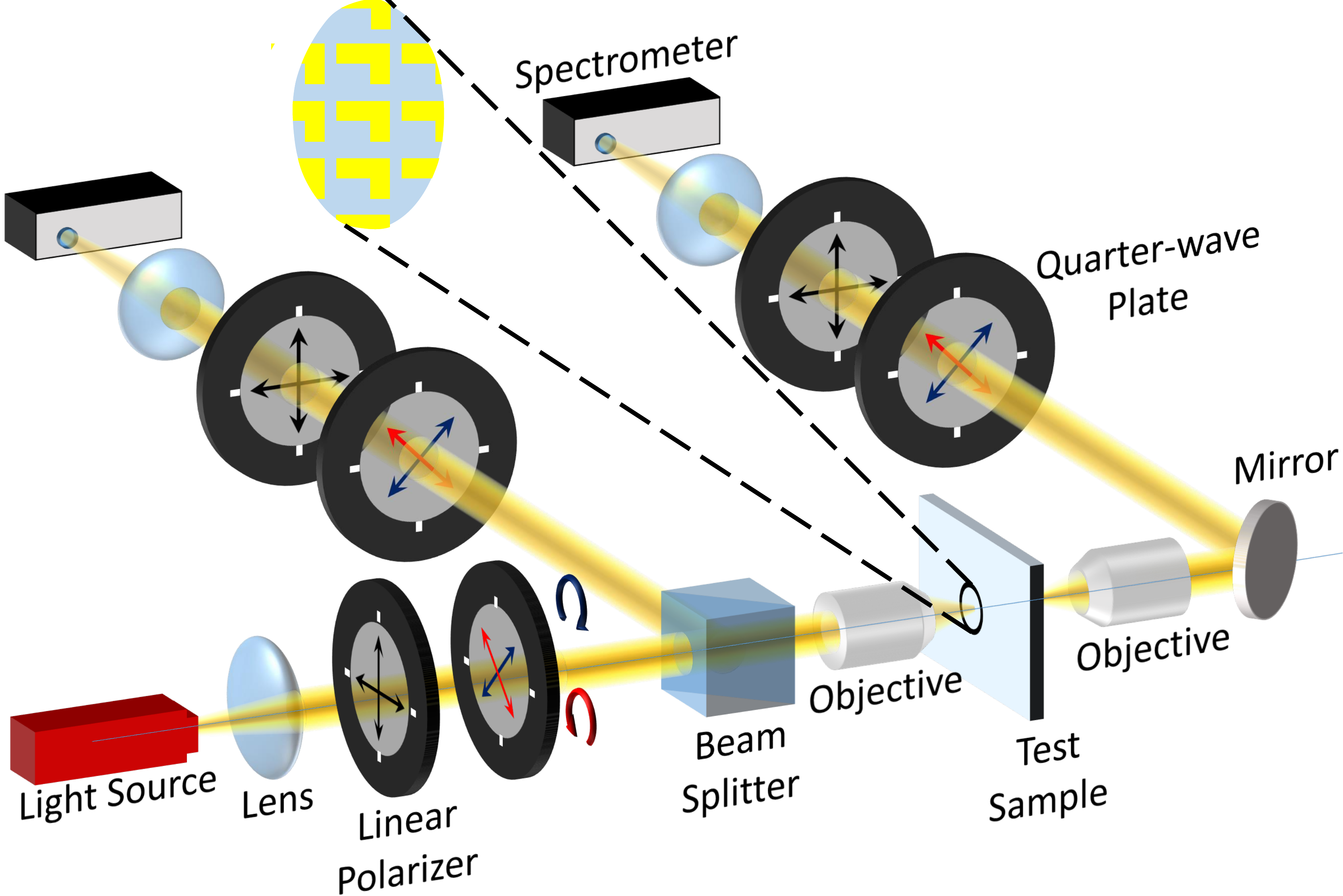, width=0.92\columnwidth}}
  \caption{In a typical optical set-up,   light passes through a lens and then is polarized using a linear polarizer. A beam splitter is required to enable the simultaneous  measurement of the reflection and transmission spectra. In the case of a CP basis, a quarter-wave plate polarizer is additionally required in the incident path. In the transmission (reflection) path, a linear and/or a quarter-wave plate analyzer is needed, and a spectrometer reads the received light power after passing through a lens.}\label{fig_opt_set}
\end{figure}

In order to fully characterize a reciprocal bianisotropic metasurface, from Eqs.~(S.5)--(S.12) in an LP basis  or from Eqs.~(S.15)--(S.22) in a CP basis, there are a maximum of $10$ in-plane polarizability components to be determined. As it is clear [see, e.g., Sec.~S--7 in Supplementary Material], Eqs.~(S.1) and~(S.2) with the help of Eqs.~(\ref{chirality2D_Rev1:our_collective}) are enough to identify all polarizability components if both amplitudes and phases of the reflected and transmitted fields are known for illumination from both directions (from the top and bottom side of the metasurface). However, as mentioned this is not always the case, especially when the information about the reflection and transmission phases are not easily and accurately measurable. Therefore, in practice measurements are often limited to amplitude-only data of the reflection and transmission coefficients, and often such data is obtained using only one direction of illumination, i.e., from one side of the metasurface  only.
We stress that in realistic optical characterization setups, measuring spectroscopy data with both illumination directions is not always possible. For example, some bio-samples that are weakly attached to one side of the supporting substrate must be always faced up. In a microscopy system, full reflection/transmission for both top and bottom illumination is often not possible~\cite{JinMoh1}. Also, in spectroscopic instruments such as in ellipsometry, the incident port is fixed and the sample slide must be mounted at the designated mechanism and the sample side must face out to the incident beam and only reflection is measured~\cite{JinMoh2}. In short, there are many cases where the illumination is convenient only from one particular side of the sample. As an example, a typical experimental optical set-up for measuring the reflection and transmission is shown in Fig.~\ref{fig_opt_set}. 

{\color{black} We emphasize once more that classification and some symmetry properties of planar structures and metasurfaces have been solely done  based on complex  reflection/transmission parameters  to some extent and in part (see Refs.~\cite{Albooyeh,kim2021chiroptical,kruk2015polarization,arteaga2014useful,li2013chiral,isik2009analysis,maslovski2009symmetry,bai2007optical,tretyakov2003analytical,achouri2019angular,achouri2021fundamental,menzel2010advanced}).} In the following, however, we investigate metasurface properties (based on the introduced classes in the Sec.~\ref{ana_st}) from {\it amplitude-only} data of the scattering parameters (reflection and transmission coefficients), and considering {\it one-side illumination} only. 

{\footnotesize
\begin{table*}[]
\centering
\caption{Distinguishing metasurface classes from amplitude-only data of reflection/transmission.}
\label{Tab2}
{%
\begin{tabular}{|c|c|c|c|c|c|c|}
\hline
\multirow{2}{*}{} & \multicolumn{2}{c|}{condition} & \multicolumn{2}{c|}{expression} & \multirow{2}{*}{} & \multirow{2}{*}{} \\ \cline{2-5}
  &  &  &  &  & &  \\
 & LP & CP & LP & CP & amplitude-only & one-side illumination \\ 
   &  &  &  &  & &  \\
\hline
  &  &  &  &  & &  \\
anisotropic &  $|t_{ii}|\neq |t_{jj}|$ & $|t_{\rm RL/LR}|\neq 0$ & Eqs.~(S.5),~(S.6) & Eqs.~(S.21),~(S.22) & \cmark & \cmark \\
 &  &  &  &  & &  \\
\hline
 &  &  &  &  & &  \\
chiral & $|t_{ij}| \neq |t_{ji}|$ & $|t_{\rm RR}|\neq|t_{\rm LL}|$ & Eqs.~\r{chiral_lp},~\r{chiral_lp_iso} & Eq.~(\ref{chirality2D_Rev1:CDt_co}) & \cmark & \cmark \\ 
 &  &  &  &  & &  \\
\hline
 &  &  &  &  & &  \\
omega & $|r^+_{ii}|\neq|r^-_{ii}|$ & $|r^+_{\rm RL/LR}|\neq|r^-_{\rm RL/LR}|$ & Eq.~\r{omega} & Eq.~\r{omega_cp} & \cmark & \xmark \\ 
 &  &  &  &  & &  \\
\hline
\end{tabular}%
}
\end{table*}
}

\section{Distinguishing metasurface classes from amplitude-only reflection and transmission data}\label{disting}
The problem is to find out to what extent we can identify or characterize the properties of a reciprocal bianisotropic metasurface (in terms of the aforementioned classes) from amplitude-only data of reflection and transmission coefficients either in an LP basis or in a CP basis and for one-side-only illumination.   We note that the analysis and synthesis of non-reciprocal metasurfaces would be more complicated than what we study here and it would require a separate investigation. In the following subsections A and B, we only consider illumination from the top side (i.e., from the $z>0$ half-space, with the incident plane wave propagating along the $-z$ direction) unless otherwise specified, hence, for brevity we drop the superscript sign $\pm$ in the reflection and transmission coefficients $r$ and $t$.

\subsection{Anisotropic metasurfaces}\label{aniso_prop}
Let us first identify whether a metasurface is anisotropic (i.e., it is made of anisotropic elements or the lattice is anisotropic). It is seen from Eqs.~(S.5) and~(S.6) that if $|t_{xx}|\neq |t_{yy}|$, then, the metasurface is anisotropic. {\color{black}A similar conclusion is observed in a CP basis when $|t_{\rm RL}|\neq0$ and/or $|t_{\rm LR}|\neq 0$.} {\color{black}These are necessary but not   sufficient conditions since, e.g., if $\eta\alpha_{ii}^{\rm ee}=\alpha_{ii}^{\rm mm}/\eta=0$ and $\eta\alpha_{xy}^{\rm ee}=\alpha_{yx}^{\rm mm}/\eta$, then, anisotropy is not recognizable using the above conditions and one needs to refer to Eqs.~(S.5)--(S.12) in an LP basis or Eqs.~(S.15)--(S.22) in a CP basis to correctly determine whether a metasurface is anisotropic.}

If a metasurface is impenetrable there is no transmission, and the identification of anisotropy is not straightforward from amplitude-only data in either of the LP or CP bases in the general case. The reason is explained as following. In the LP basis and referring to the co-polarized reflection coefficients $|r_{xx/yy}|$, if the metasurface is anisotropic, then, $\alpha_{xx}^{\rm ee,mm}\neq \alpha_{yy}^{\rm ee,mm}$. However, due to the presence of omega properties $\alpha_{xy,yx}^{\rm em,em}$ in the LP reflection coefficients, anisotropic and omega properties are mixed up and indistinguishable. A similar situation is valid for anisotropic and chiral properties in the cross-polarized reflection coefficients $|r_{xy/yx}^{\pm}|$. In the CP basis, the situation is even more complicated since in the co-component reflection coefficients $|r_{\rm RR/LL}^{\pm}|$, both omega $\alpha_{xy,yx}^{\rm em,em}$ and chiral $\alpha_{xx,yy}^{\rm em,em}$ properties are present together with the anisotropic ones.

\subsection{Chiral metasurfaces: inversional asymmetry}\label{chiral_prop}
In an LP basis, and from amplitude-only data from one-side-only illumination, chirality can only be probed from the amplitude of the cross-polarized transmission coefficients and only {\color{black}if $|t_{xy}| \neq |t_{yx}|$. Under such condition, the metasurface is chiral since from Eqs.~(S.7) and~(S.8)  \e
|t_{xy}|^2-|t_{yx}|^2={\omega^2\over S^2}\Re \{ {\color{black}(\eta{\alpha}^{\textrm{ee}}_{xy}-{{\alpha}^{\textrm{mm}}_{yx}\over \eta})^\ast}{\color{black}({\alpha}^{\textrm{em}}_{xx}+{\alpha}^{\textrm{em}}_{yy})} \}.\l{chiral_lp}
\f
Here the superscript ``$^\ast$'' denotes the complex conjugation. 
It is worth mentioning that the larger values of $|t_{xy}|^2-|t_{yx}|^2$ in Eq.~\r{chiral_lp} does not necessarily imply stronger chirality [i.e., larger ${\color{black}({\alpha}^{\textrm{em}}_{xx}+{\alpha}^{\textrm{em}}_{yy})}$] since a large difference in transmission could also be due to the multiplying term ${\color{black}(\eta{\alpha}^{\textrm{ee}}_{xy}-{{\alpha}^{\textrm{mm}}_{yx}/ \eta})}$. Moreover, if condition ${\color{black}\eta{\alpha}^{\textrm{ee}}_{xy}-{{\alpha}^{\textrm{mm}}_{yx}/ \eta}}=0$ is satisfied (the isotropic case ${\alpha}^{\textrm{ee}}_{xy}={\alpha}^{\textrm{mm}}_{yx}=0$ is a particular case), Eq.~\r{chiral_lp} vanishes and chirality cannot be probed with this equation. Thus, a nonzero value of Eq.~\r{chiral_lp} is only a sufficient and not necessary condition for chirality detection. Note that it is not trivial for anisotropic metasurfaces to satisfy condition ${\color{black}\eta{\alpha}^{\textrm{ee}}_{xy}-{{\alpha}^{\textrm{mm}}_{yx}/ \eta}}=0$, hence, in most cases, Eq.~\r{chiral_lp} is a proper condition to verify if an anisotropic metasurface is also chiral. If a metasurface is isotropic in its plane (i.e, its constituents are isotropic in the $x-y$ plane), which implies that $|t_{ii}|=|t_{jj}|$, then chirality is easily recognized if either one of the LP cross-polarized transmission amplitudes are nonzero, i.e., $|t_{ij}|\neq 0$, and it is obtained as
\begin{equation}\l{chiral_lp_iso}
|t_{ij}|={\omega\over{S}} |{\color{black}{\alpha}^{\textrm{em}}}|.
\end{equation}
}

In a CP basis, examination of chirality is much easier since it is only required to observe that $|t_{\rm RR}| \neq |t_{\rm LL}|$ no matter whether the metasurface is isotropic or not in its plane. This is one reason why the CP basis is largely used for detection of chiral samples. In this case, from Eqs.~(S.19) and (S.20) we obtain


\begin{widetext}
\begin{eqnarray}
\Delta T_{\rm co} &=&  2{\omega\over S}~\Re\left[\left(1-{j\omega\over{4S}}\left[\underbrace{\eta \left({\alpha}^{\textrm{ee}}_{xx}+{\alpha}^{\textrm{ee}}_{yy}\right)+\left({\alpha}^{\textrm{mm}}_{xx}+{\alpha}^{\textrm{mm}}_{yy}\right)/\eta}_{\rm electric+magnetic}\right]\right)^\ast {\color{black}\underbrace{\left({\alpha}^{\textrm{em}}_{xx}+{\alpha}^{\textrm{em}}_{yy}\right)}_{\rm chirality}}\right], \label{chirality2D_Rev1:CDt_co}
\end{eqnarray}
\end{widetext}

Here, $\Delta T_{\rm co}=|t_{\rm RR}|^2-|t_{\rm LL}|^2$. For the case with illumination from the bottom side of the metasurface, we obtain the same expression as in Eq.~(\ref{chirality2D_Rev1:CDt_co}). It is important to note that metasurface chirality cannot be recognized from the reflection amplitudes in either of the LP or CP basis. This is because the chirality term is always summed up with an anisotropy term in a similar fashion (i.e., same sign difference) in either CP or LP basis. Note that strictly speaking a nonzero value obtained by Eq.~(\ref{chirality2D_Rev1:CDt_co}) is only a sufficient but not necessary  condition for chirality detection because when $\left[\eta \left({\alpha}^{\textrm{ee}}_{xx}+{\alpha}^{\textrm{ee}}_{yy}\right)+\left({\alpha}^{\textrm{mm}}_{xx}+{\alpha}^{\textrm{mm}}_{yy}\right)/\eta\right]={-j4S/{\omega}}$, chirality is not recognizable from this equation. However, in most practical scenarios this latter equality is not satisfied, especially because electric polarizabilities are often stronger than magnetic ones, especially in optics, and Eq.~(\ref{chirality2D_Rev1:CDt_co}) is a proper criterion to determine whether a metasurface is chiral or not.

Another important consideration [see Eqs.~(S.7) and~(S.8)] is that if the metasurface is chiral, one clearly observes asymmetric behavior for the cross-component transmisstion coefficients for opposite illumination directions in an LP basis, i.e., $|t_{xy/yx}^+|\neq |t_{xy/yx}^-|$. Such a behavior has been experimentally studied and observed in metasurfaces~\cite{menzel}.

\subsection{Omega metasurfaces: translational asymmetry}\label{omega_prop}

{\color{black} We investigate here what kind of plane wave polarizations and reflection/transmission observations carry information about the EM omega  properties of the metasurface, i.e., when it is possible to retrieve non-vanishing $\alpha_{\rm em}^{ij}$ where $i\neq j$.} It is obvious from Eqs.~(S.5)--(S.12) in the LP basis and Eqs.~(S.15)--(S.22) in the CP basis, that omega properties cannot be identified from any of the transmission data. It only manifests in {\it reflection} data. Furthermore, this property is never recognizable if one probes only the reflection data from one-side illumination, either in LP nor in CP basis. Therefore, the identification of omega properties remains a challenge if an optical set-up is designed for measurement with one-side illumination only. Indeed, if one is able to provide two separate incident fields which illuminate the metasurface form both sides, then, the identification of omega property is easily possible in either CP or LP basis from reflection-only amplitudes since from Eqs.~(S.9) and~(S.10) one has
\e
|r^+_{ii}|^2-|r^-_{ii}|^2=- 2{\omega^2\over S^2}\Re \{ {\color{black}(\eta{\alpha}^{\textrm{ee}}_{ii}-{{\alpha}^{\textrm{mm}}_{jj}\over \eta})^\ast}{\color{black}{\alpha}^{\textrm{em}}_{ij}} \},\l{omega}
\f
in an LP basis, where the ``$^+$'' or ``$^-$'' superscript sign  represents the respective illumination from the top or bottom side, and $i,j=x~{\rm or}~y$. Note that if Eq.~\r{omega} equals zero, then, either or both terms $\eta{\alpha}^{\textrm{ee}}_{ii}-{{\alpha}^{\textrm{mm}}_{jj}/ \eta}$ and  ${\color{black}  {\alpha}^{\textrm{em}}_{ij}}$ are vanishing. In case of $(\eta{\alpha}^{\textrm{ee}}_{ii}-{{\alpha}^{\textrm{mm}}_{jj}/ \eta})=0$ and ${\color{black}  {\alpha}^{\textrm{em}}_{ij}}\neq 0$, the matasurface reveals its omega property with $|r_{ii}|= {\omega\over{S}}|{\color{black}  {\alpha}^{\textrm{em}}_{ij}}|$. Instead, when   $(\eta{\alpha}^{\textrm{ee}}_{ii}-{{\alpha}^{\textrm{mm}}_{jj}/ \eta})\neq 0$ and ${\color{black}  {\alpha}^{\textrm{em}}_{ij}}= 0$ the metasurface does not contain omega properties. Therefore, a nonzero value of Eq.~\r{omega} is a sufficient and not necessary condition in recognizing a metasurface omega property. 

An analogous result is obtained from the cross-polarized reflection coefficients $r_{\rm RL,LR}$ in CP basis from ether of Eqs.~(S.17) or~(S.18) leading to
\begin{widetext}
\begin{equation}
\begin{array}{l}
\displaystyle
|r_{\rm LR}^{+}|^2-|r_{\rm LR}^{-}|^2=-2{\omega^2\over{S^2}}\Re\left(\left[\eta \left({\alpha}^{\textrm{ee}}_{xx}+{\alpha}^{\textrm{ee}}_{yy}\right)-{1\over \eta} \left({\alpha}^{\textrm{mm}}_{xx}+{\alpha}^{\textrm{mm}}_{yy}\right)\right]^\ast{\color{black} \left({\alpha}^{\textrm{em}}_{xy}-{\alpha}^{\textrm{em}}_{yx}\right)}\right).
\end{array}
\l{omega_cp}
\end{equation}
\end{widetext}

Similarly to the previous cases, a nonzero value of Eq.~\r{omega_cp} is only a sufficient and not necessary  condition for the detection of omega properties in a metasurface since if $\eta \left({\alpha}^{\textrm{ee}}_{xx}+{\alpha}^{\textrm{ee}}_{yy}\right)={1\over \eta} \left({\alpha}^{\textrm{mm}}_{xx}+{\alpha}^{\textrm{mm}}_{yy}\right)$, then the right-hand side of this equation vanishes while the metasurface may yet possess omega property due to the non-zero  ${\alpha}^{\textrm{em}}_{xy}-{\alpha}^{\textrm{em}}_{yx}$ term. Such a situation happens for instance in case of metasurfaces with total absorption~\cite{pozar,AlbooyehOE}. {\color{black} Note that if we allow the illumination to be off normal, we would be able to retrieve omega properties from one side illumination only. In that case we require reflection data for different incident angles, and moreover, the retrieval procedure seems to be more complicated compared to the case of normal illumination.}


\section{Summary of metasurface classes}
In Table~\ref{Tab2} and below with bullet points we summarize how to distinguish metasurface classes using amplitude-only data of reflection/transmission coefficients and for illumination from one side only.

\begin{itemize}

\item Anisotropy is determined from the co- (cross-)pol {\it transmission} amplitudes in LP (CP) basis. It is not easily recognizable from reflection data.

\item  Chiral metasurfaces are not recognized from anisotropic ones via {\it reflection} amplitudes since the parameters corresponding to both chirality and anisotropy are summed up similarly in the reflection coefficients. However, they are recognizable from the {\it transmission} amplitudes of two orthogonal polarization scenarios of the incident waves.


\item Omega properties cannot be recognized from data coming from one side illumination only and  one set of amplitude-only data of normal incidence. They are neither recognized from the transmission coefficients. To identify whether a metasurface contains omega properties one requires to have {\it reflection} amplitude data from opposite illumination directions in a normal excitation scenario, or two (or even more) sets of {\it reflection} amplitude data from one side illumination  for different illumination angles (not shown in the equations of this study, see Ref.~\cite{Albooyeh} for a more elaborated discussions).


\end{itemize}

\section{CD experiment and metasurface properties}\label{CD_exp}
Two well-known quantities which are used in characterization of chiral media are CD and the dissymmetry factor (or  $g$-factor)~\cite{schwanecke,Ambilino,Barron,mousaviL}. They are two analogous criteria to measure the difference between {\it extinction} of a material under two illuminations with opposite optical helicity (i.e., EM field chirality), typically using both right-hand CP (RCP) and left-hand CP (LCP) light~\cite[Sec.11]{Fasman},~\cite{lindell,Gvozdev, Tang1, Mina1}. In practice, the difference between the {\it transmittance} spectra for an RCP and an LCP light illumination are used in determination of the sample's chirality~\cite{Frank,Esposito}. Indeed, it is widely accepted that the difference between the transmittance spectra for LCP and RCP light is due to material chirality. However this speculation is only correct for {\it isotropic} chiral bulk media~\cite{lindell} and cannot be always concluded for a general anisotropic media with all properties involved. Note that bulk media (including molecules in solution) with fully randomly arranged inclusions are also considered as isotropic since there is no statistical directional preferences. For metasurfaces extra precautions must be taken. It is proven that for a periodic structure the normalized extinction power $P_{\rm ext}$ is related to the transmission coefficient $t$ as $P_{\rm ext}=2\Re{(1-t)}P_{\rm inc}$, where $P_{\rm inc}$ is the incident power~\cite{Gustafsson, CampioneS}. As it is clear from Eqs.~(S.19)--(S.22), for an isotropic metasurface $|t_{\rm RL}|=|t_{\rm LR}|=0$, and if $\left[\eta \left({\alpha}^{\textrm{ee}}_{xx}+{\alpha}^{\textrm{ee}}_{yy}\right)+\left({\alpha}^{\textrm{mm}}_{xx}+{\alpha}^{\textrm{mm}}_{yy}\right)/\eta\right]\neq{-j4S/{\omega}}$ (which is true in most situations) then Eq.~(\ref{chirality2D_Rev1:CDt_co}) that is the difference between  co-polarized transmittances for RCP and LCP light  is a measure of the metasurface chirality. However, for an {\it anisotropic} metasurface the situation is not straightforward and needs to be investigated as follows.

Here, we consider CD looking at the transmission spectra since it is practically impossible to separate the effect of chirality from anisotropy by looking at reflection coefficients, as discussed in Sec.~\ref{disting}.  As widely used for the transmission spectra, the {\it normalized} CD for a sample is defined as the difference between the total transmittance (square of the transmission amplitude) of LCP and RCP waves divided by the arithmetic average of the transmittance for these two illuminations, i.e.,~\cite{Frank,Esposito}
\e
\displaystyle
{\rm CD_{tot}}={{|t_{\rm LL}|^2+|t_{\rm RL}|^2-{|t_{\rm RR}|^2-|t_{\rm LR}|^2}}\over {{1\over 2}\left( {|t_{\rm LL}|^2+|t_{\rm RL}|^2}+{|t_{\rm RR}|^2+|t_{\rm LR}|^2}\right) }}. \label{CDtra}
\f 
We emphasize that the above definition for CD is not the conventional one (as in, e.g., Ref.~\cite[Sec.~1.5.1]{Ambilino} and Ref.~\cite[Sec. 3.4.6]{Barron_mol})  since here, for practical purposes, transmittance is used rather than the absorptance. Moreover, the division by the arithmetic average in Eq.~(\ref{CDtra}) is normally used also in the definition of the dissymmetry factor g ($g$-factor). {\color{black} Alternative definition of CD based on absorptance can be found in Sec.~S--9 of Supplementary Material. Upon that definition, chirality is only detectable  if the metasurface is isotropic in its plane, and moreover, if it does not contain omega coupling. A similar conclusion is given in Ref.~\cite[Sec. 3, see discussions after Eq. (3.4.40)]{Barron_mol}. Importantly, in the general anisotropic case, CD based on absorptance cannot be used to determine the presence of chirality.}

Next, we decompose the contributions of the co- and cross-polarized components of transmittance in the CD defined by Eq.~(\ref{CDtra}). This is achieved by using polarizers in experimental set-ups. That is, we consider the difference between the co-polarized transmittances as $\Delta T_{\rm co}=|t_{\rm LL}|^2-|t_{\rm RR}|^2$, and the cross-polarizedtransmittances as $\Delta T_{\rm cr}=|t_{\rm RL}|^2-|t_{\rm LR}|^2$. The expression for $\Delta T_{\rm co}$  is already given in Eq.~(\ref{chirality2D_Rev1:CDt_co})
and $\Delta T_{\rm cr}$ is obtained by using Eqs.~(S.21) as

\begin{widetext}
\begin{eqnarray}
\Delta T_{\rm cr} &=& {\omega\over S}~\Im\left[{\omega\over 2S}\left(\underbrace{\eta \left({\alpha}^{\textrm{ee}}_{xx}-{\alpha}^{\textrm{ee}}_{yy}\right)-\left({\alpha}^{\textrm{mm}}_{xx}-{\alpha}^{\textrm{mm}}_{yy}\right)/\eta}_{\rm anisotropy}\right)^\ast {\color{black}\underbrace{\left(\eta{\alpha}^{\textrm{ee}}_{xy}-{\alpha}^{\textrm{mm}}_{yx}/\eta\right)}_{\rm anisotropy}}\right].
\label{chirality2D_Rev1:CDt_xs}
\end{eqnarray}
\end{widetext}
 As it is clear from Eqs.~(\ref{chirality2D_Rev1:CDt_co}) and~(\ref{chirality2D_Rev1:CDt_xs}), $\Delta T_{\rm co}$ describes both chirality (if ${\alpha}^{\textrm{em}}_{xx}\neq 0$ or ${\alpha}^{\textrm{em}}_{yy}\neq 0$) and anisotropy (if ${\alpha}^{\textrm{ee}}_{xx}\neq {\alpha}^{\textrm{ee}}_{yy}$ or ${\alpha}^{\textrm{mm}}_{xx}\neq {\alpha}^{\textrm{mm}}_{yy}$) effects whereas $\Delta T_{\rm cr}$ describes only the anisotropy properties of the metasurface. That is, if $\Delta T_{\rm co}=0$ (which in most cases implies no chirality as discussed in Sec.~\ref{chiral_prop}), there is still a possibility to observe a considerable CD if $\Delta T_{\rm cr}\neq 0$, which means anisotropy. Therefore, with the goal of distinguishing anisotropy from chirality effects, we decompose Eq.~(\ref{CDtra}) into two terms and introduce two CDs; one with the contribution of only co-polarized transmission data, i.e.,  
\e
\displaystyle
{\rm CD_{chi}}={{|t_{\rm LL}|^2-|t_{\rm RR}|^2}\over {{1\over 2}\left( {|t_{\rm LL}|^2+|t_{\rm RL}|^2}+{|t_{\rm RR}|^2+|t_{\rm LR}|^2}\right) }}. \l{CDchi}
\f 
which we call CD from metasurface chirality, whereas the other with the contribution of cross-polarized transmission data, i.e.,
\e
\displaystyle
{\rm CD_{ani}}={{|t_{\rm RL}|^2-|t_{\rm LR}|^2}\over {{1\over 2}\left( {|t_{\rm LL}|^2+|t_{\rm RL}|^2}+{|t_{\rm RR}|^2+|t_{\rm LR}|^2}\right) }}, \l{CDani}
\f 
which we call CD from metasurface anisotropy. Note that these definitions do not carry any information regarding the contribution of other material properties such as omega. However, they clearly determine whether a metasurface possesses chirality, unless  $\left[\eta \left({\alpha}^{\textrm{ee}}_{xx}+{\alpha}^{\textrm{ee}}_{yy}\right)+\left({\alpha}^{\textrm{mm}}_{xx}+{\alpha}^{\textrm{mm}}_{yy}\right)/\eta\right]={-j4S/{\omega}}$. That is, if  ${\rm CD_{chi}}=0$, the metasurface is clearly not chiral (unless $\left[\eta \left({\alpha}^{\textrm{ee}}_{xx}+{\alpha}^{\textrm{ee}}_{yy}\right)+\left({\alpha}^{\textrm{mm}}_{xx}+{\alpha}^{\textrm{mm}}_{yy}\right)/\eta\right]={-j4S/{\omega}}$ which may yet be chiral and it is a rare situation). A nonzero ${\rm CD_{ani}}$ proves that the metasurface is anisotropic.  Moreover, since anisotropy is included in both ${\rm CD_{chi}}$ and ${\rm CD_{ani}}$ definitions, one cannot necessarily conclude anything about the metasurface anisotropy if only ${\rm CD_{ani}}=0$. That is, a metasurface can be anisotropic while ${\rm CD_{ani}}=0$. This case happens if $\alpha^{\rm ee,mm}_{xy}=0$ and $\alpha^{\rm ee,mm}_{xx}\neq\alpha^{\rm ee,mm}_{yy}$. Therefore, to have a complete judgment about metasurface anisotropy, one requires to perform an analysis in an LP basis as discussed before and as it will be demonstrated in Sec.~\ref{chirality2D_Rev1:L-shaped2}. As a final note, we mention that values of $\rm CD_{ chi}$ and ${\rm CD_{ani}}$ do not provide a direct information about chirality or anisotropy strengths since they involve various other terms.

It is now beneficial to study some special metasurface cases that provide anisotropy and/or chirality characteristics to illustrate how metasurfaces are characterized based on our rigorous analysis. 

\section{Illustrative examples}

We consider several metasurface examples; some of them have been previously analyzed in the literature and it has been experimentally proven that they exhibit CD. However, as discussed earlier and shown next, CD is not exclusively provided by material's chirality in all cases. Our focus is to resolve possible confusions in the characterization of anisotropic versus chiral metasurfaces. Following similar reasoning other material characteristics such as omega properties could also be studied and can be found in the literature (e.g., Refs.~\cite{raditailoring,Albooyeh,Yazdi2,AlbooyehEqui,VAsadchy}). {\color{black}We finally state that all simulations are performed in time domain by using a finite integration technique (FIT) which is implemented in CST Studio Suite commercial software tool~\cite{Saduku,CST}.}

\begin{figure*}
\centering
   \subfigure[]{
  \epsfig{file=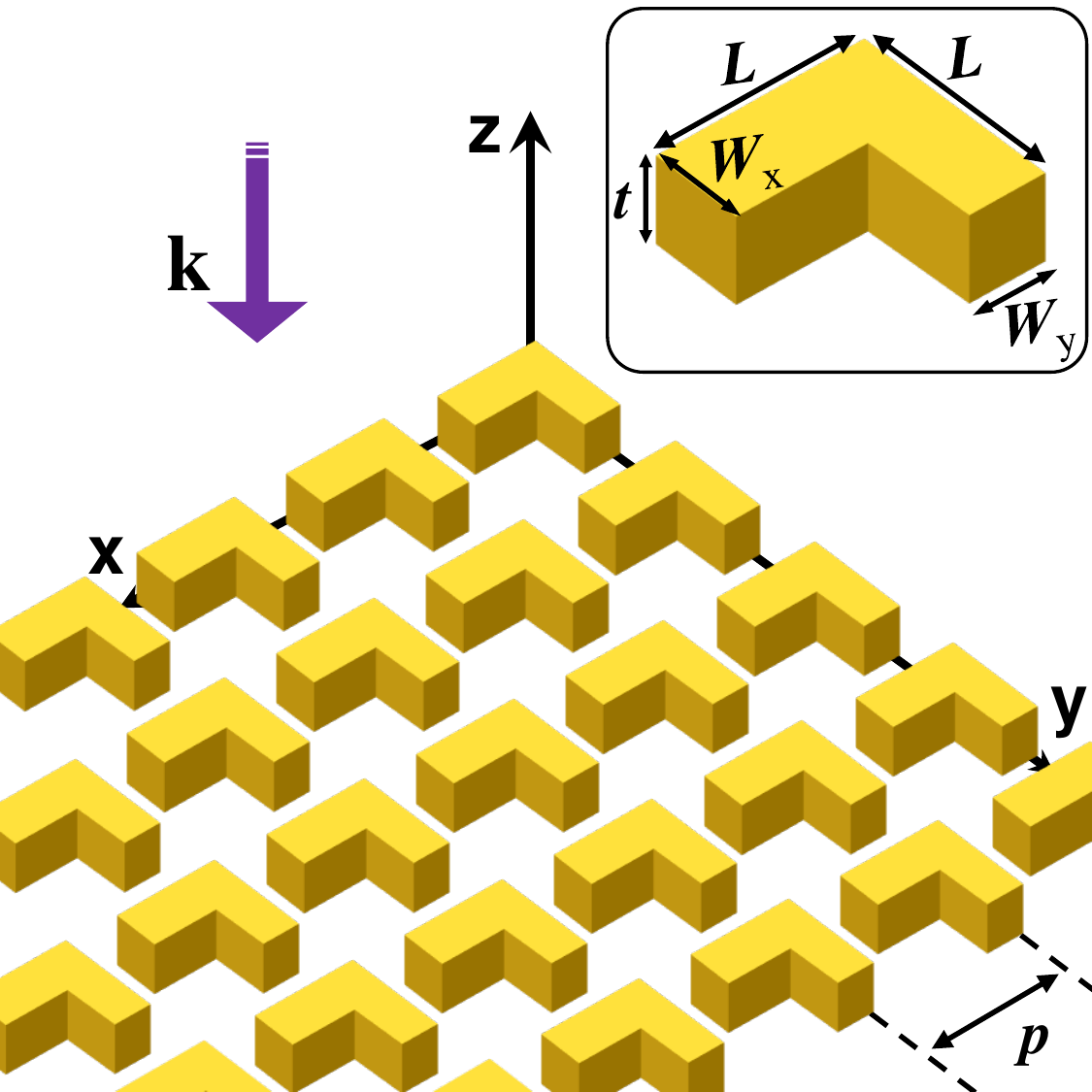, width=0.45\columnwidth}
   \label{ris:figWeimina} }
  \subfigure[]{
   \epsfig{file=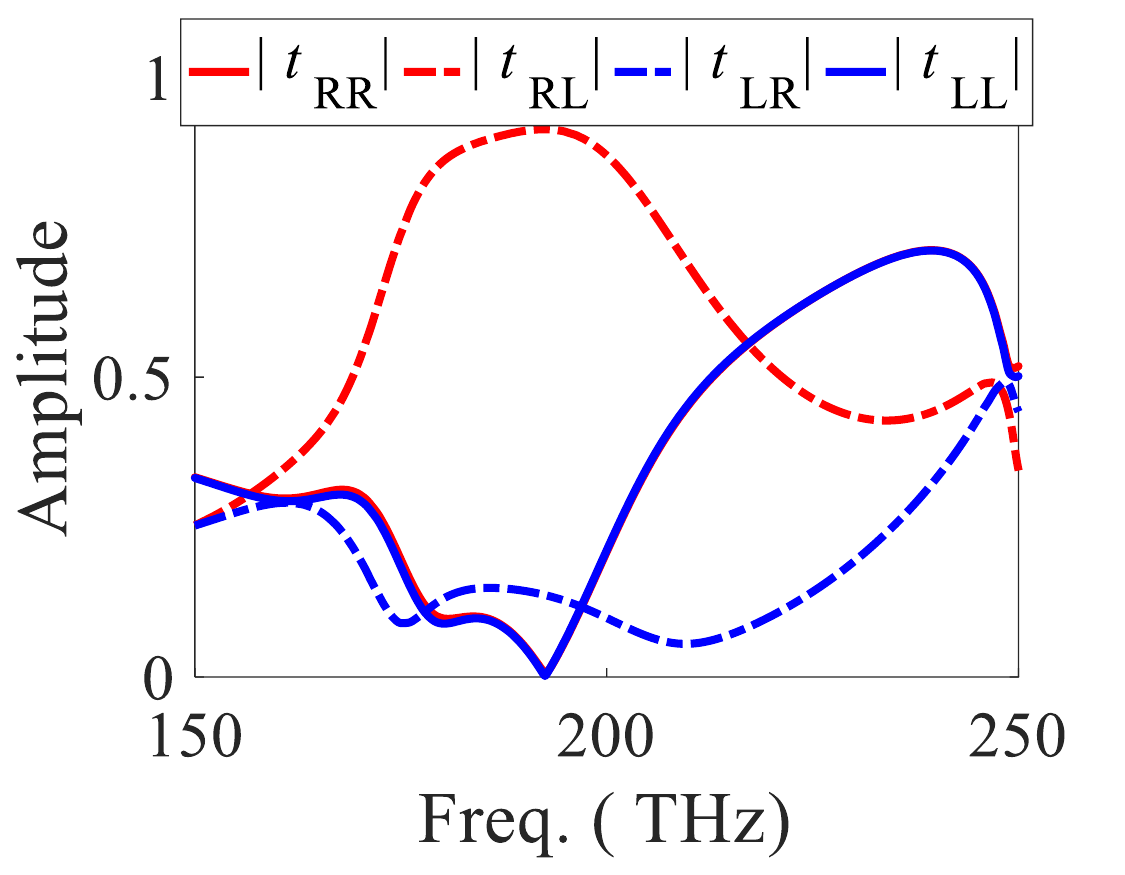, width=0.55\columnwidth}
   \label{ris:figWeiminb} }
     \subfigure[]{
   \epsfig{file=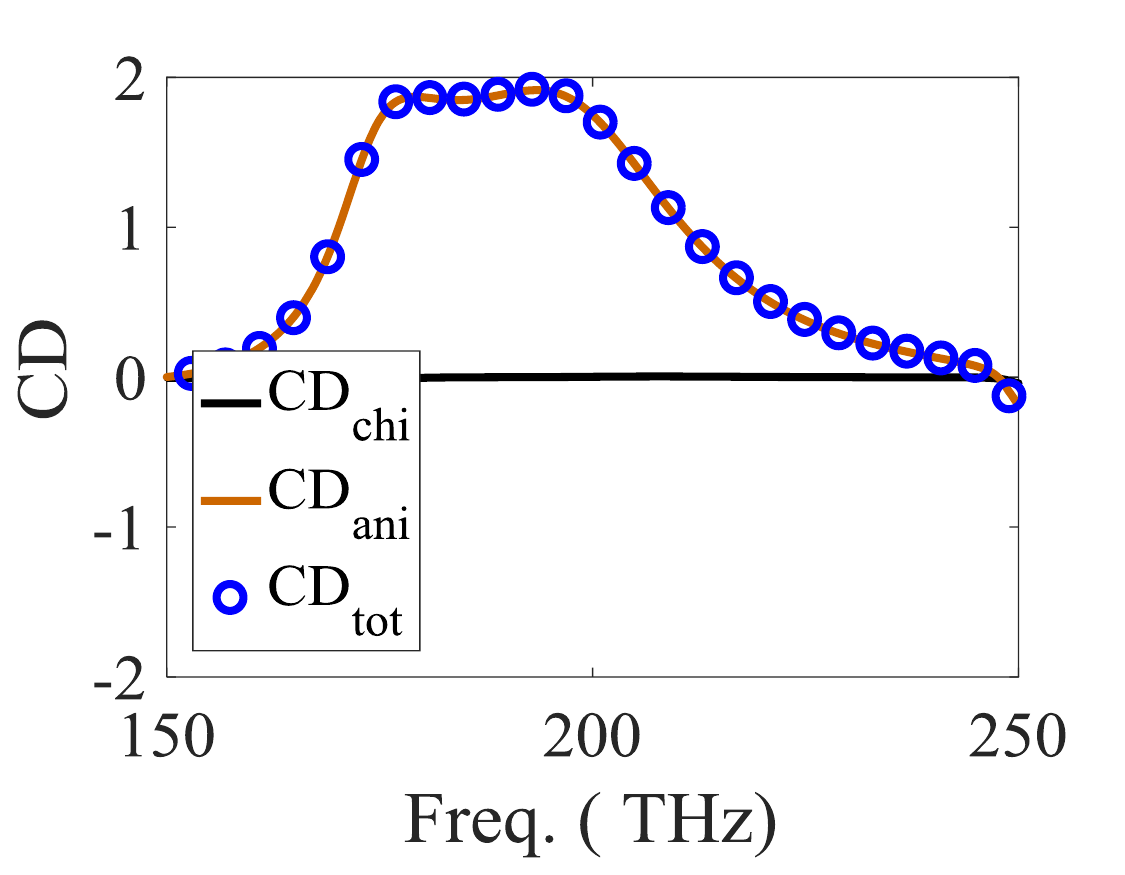, width=0.55\columnwidth}
   \label{ris:figWeiminc} }
  \caption{(a) The proposed metasurface reported in Ref.~\cite{weimin}. The metasurface is composed of a periodic array of L-shaped anisotropic particles. The gold particles are embedded in the fused silica. The parameters are, respectively, as following: $L=580$nm, $W_x=246$nm, $W_y=160$nm, $t=360$nm, and $p=730$nm. (b) The amplitude of all components of the transmission coefficients of the proposed metasurface in the CP basis. (c) The contributions of $\rm CD_{ chi}$ and ${\rm CD_{ani}}$ in the total transmission CD.}\label{ris:figWeimin}
\end{figure*}

\subsection{A metasurface of L-shpaed plasmonic particles: an achiral anisotropic metasurface} \label{anis_weimin1}

Let us start with a simple example that demonstrates a large CD at infrared. We consider a periodic metasurface of L-shaped particles as shown in Fig.~\ref{ris:figWeimina} and reported in Ref.~\cite{weimin}. We excite the metasurface with a normally incident plane wave so that we eliminate the effect of pseudochirality (extrinsic chirality) which results from oblique illumination as discussed previously. This metasurface was suggested in Ref.~\cite{weimin} to possess a chiral response, however, as we demonstrate here, it is   an  achiral metasurface and the observed CD should be attributed to the anisotropy of the scatterers rather than    chirality. To clarify this issue, we have plotted the amplitudes of different components of the transmission coefficients in Fig.~\ref{ris:figWeiminb}. As discussed in Sec.~\ref{ana_st}, the co-polarized transmission coefficients $\vert t_{\rm RR}\vert$ and $\vert t_{\rm LL}\vert$ are equal (on top of each other) which is a sign of the absence of chiral property in the metasurface as can be concluded from Eqs.~(\ref{chirality2D_Rev1:CDt_co}) and~(\ref{chirality2D_Rev1:CDt_xs}). However, the cross-polarized  transmission coefficients are very different. This demonstrates the anisotropy properties of the metasurface as discussed in Sec.~\ref{ana_st}. To further elaborate on this aspect, we have plotted total CD as well as  ${\rm CD_{chi}}$ and ${\rm CD_{ani}}$ of the proposed metasurface, as defined in Eqs.~(\ref{CDtra}), \r{CDchi}, and~\r{CDani}, respectively [see Fig~\ref{ris:figWeiminc}]. As it is obvious, the CD due to ${\rm CD_{chi}}$ is negligible compared to that due to anisotropy ${\rm CD_{ani}}$. Therefore, in this example, we attribute the total CD to the anisotropy of the metasurface and not to its chirality.\\
\begin{figure*}
\centering
   \subfigure[]{
  \epsfig{file=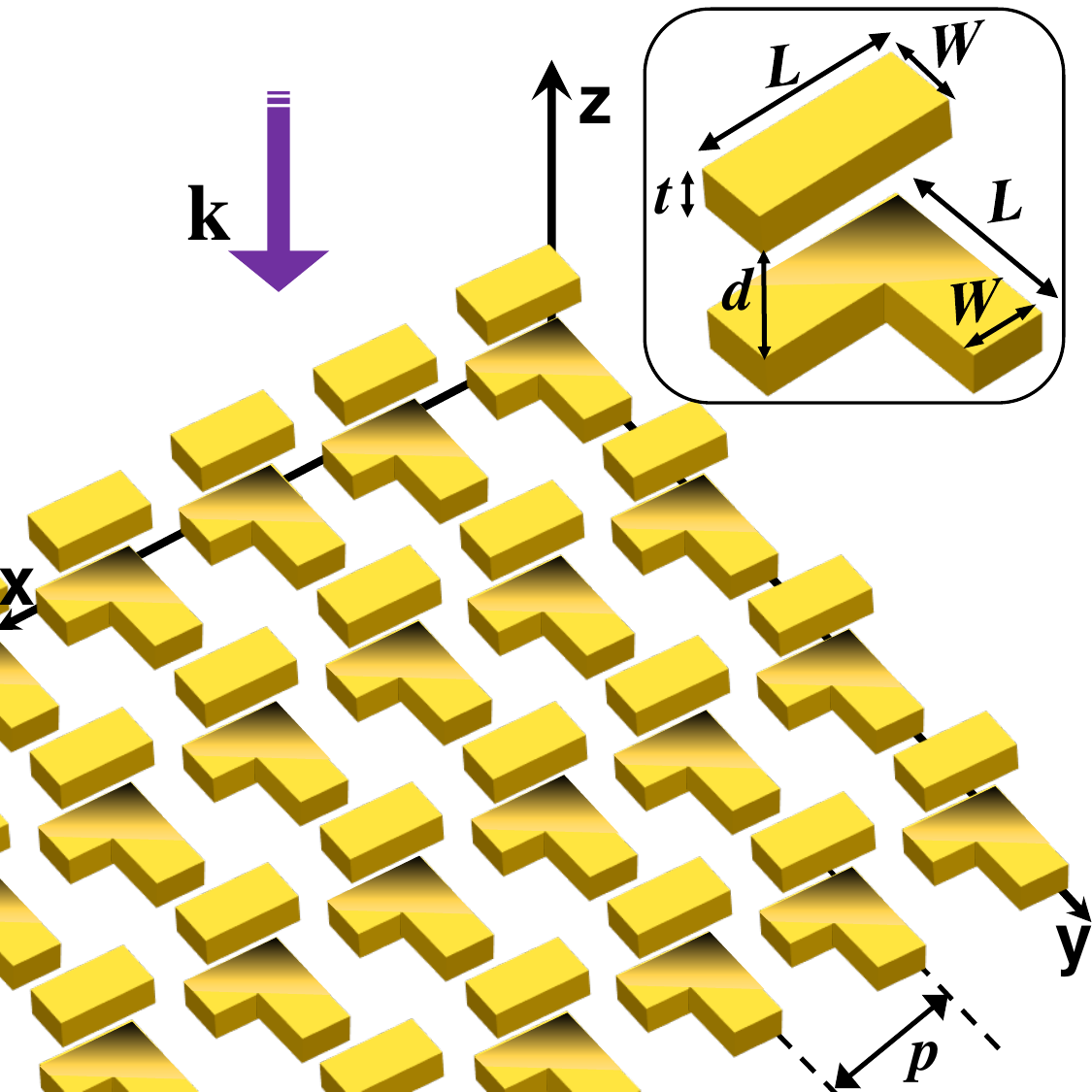, width=0.45\columnwidth}
   \label{ris:figMenzela} }
  \subfigure[]{
   \epsfig{file=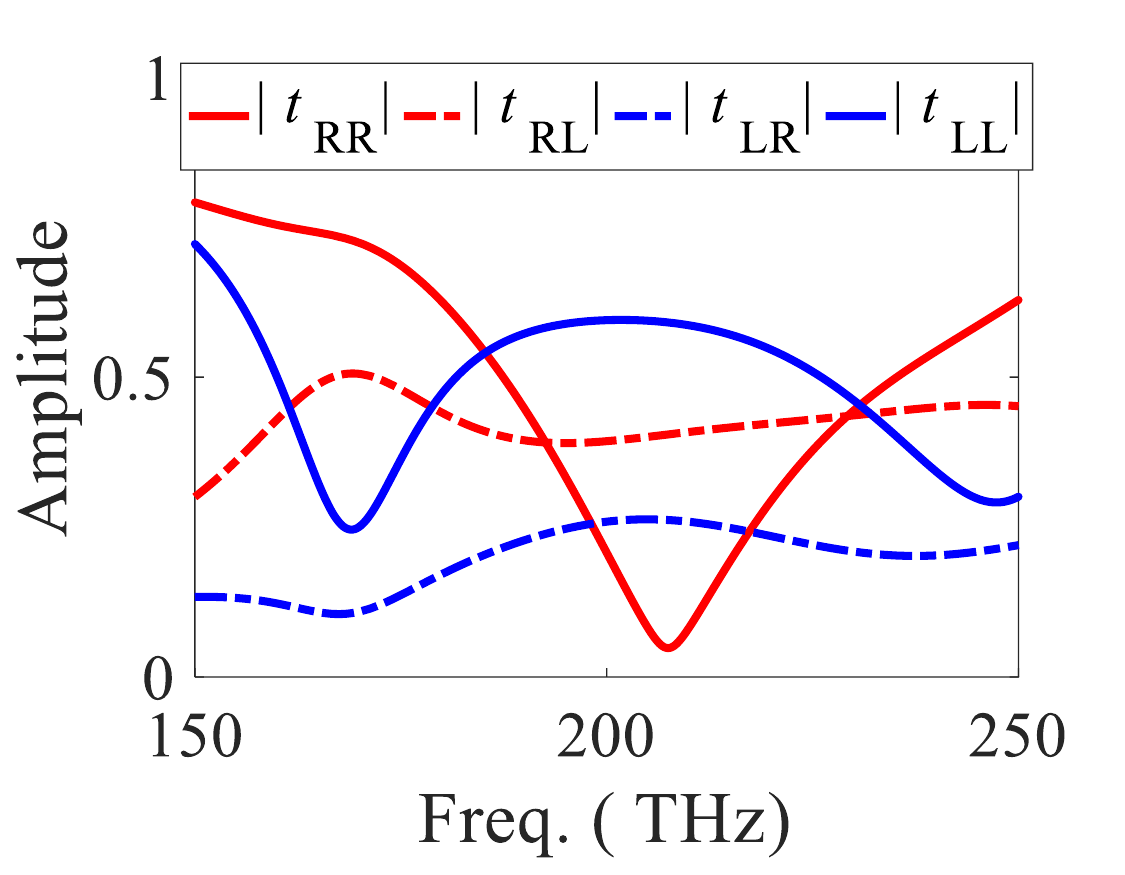, width=0.55\columnwidth}
   \label{ris:figMenzelb} }
     \subfigure[]{
   \epsfig{file=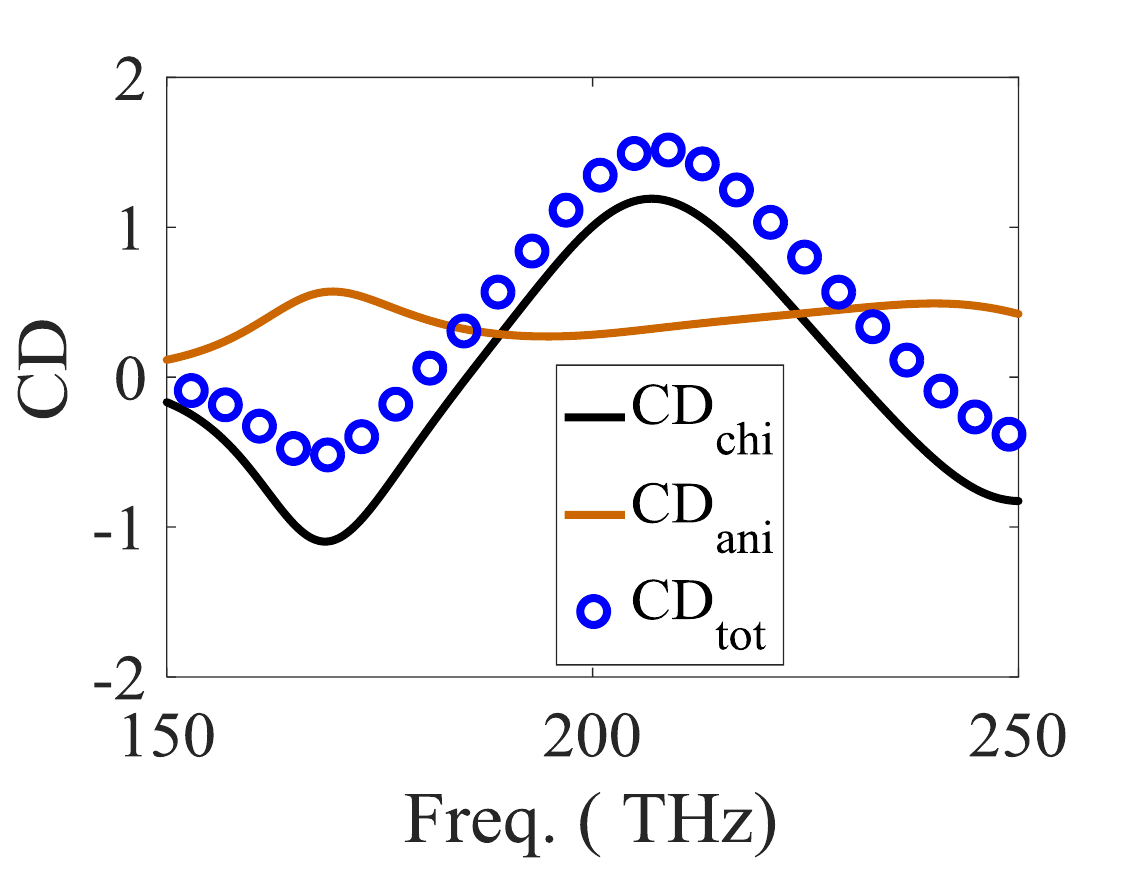, width=0.55\columnwidth}
   \label{ris:figMenzelc} }
  \caption{(a) The proposed metasurface reported in Ref.~\cite{menzel}. The metasurface is composed of two layers; strips in the top layer and L-shaped particles in the bottom layer. The gold particles are embedded in a the fused silica. Other parameters are $L=290$nm, $W=130$nm, $t=40$nm, $d=80$nm, and $p=500$nm. (b) The amplitude of all components of the transmission coefficients of the proposed metasurface in the CP basis. (c) The contributions of $\rm CD_{ chi}$ and ${\rm CD_{ani}}$ in the observed transmission CD.}\label{ris:figMenzel}
\end{figure*}

\subsection{A metasurface of L-shaped plasmonic particles loaded with plasmonic strips: an anisotropic chiral metasurface}\label{chi_menzel}
Next, we consider a metasurface which provides strong CD due to both anisotropy and chirality. This metasurface was analysed    in Ref.~\cite{menzel}. We sketched the proposed topology in Fig.~\ref{ris:figMenzela}. As it is shown in this figure, each scatterer of the metasurface is composed of a strip plasmonic particle in addition to an L-shaped particle. It may be incorrectly perceived that the observed CD is exclusively attributed to chirality of the metasurface. However, as we show here, the obtained CD is not only related to the metasurface's chirality, but also to anisotropy of metasurface. To prove this, we first plot different components of the transmission coefficients of the proposed metasurface in Fig.~\ref{ris:figMenzelb}. As it is clear from this figure, both the co- and cross-components of the transmission coefficients are not equal. Based on Eqs.~(\ref{chirality2D_Rev1:CDt_co}) and~(\ref{chirality2D_Rev1:CDt_xs}), this obviously implies the contributions of both anisotropy and chirality of the metasurface to any observed CD. To quantitatively study this issue, we have plotted   ${\rm CD_{chi}}$ and ${\rm CD_{ani}}$ of the proposed metasurface, as defined in Eqs.~\r{CDchi} and~\r{CDani}, respectively, in Fig.~\ref{ris:figMenzelc}. It is clear from this figure that the observed {\it total} CD is not exclusively due to ${\rm CD_{chi}}$ (which possesses the contributions from both chirality and anisotropy) of the proposed metasurface. That is, the part of anisotropy of the metasurface which is represented by ${\rm CD_{ani}}$ has a significant role in the total observed transmission CD. Moreover, note that metasurface anisotropy constructively adds to chirality to obtain larger CDs at some frequencies whereas its effect is destructive at some other frequencies.

The difference between this example and the example of the previous subsection is as follows. While in the previous example the structure could only be excited electrically and there was no possibility to excite it magnetically, in this second example instead the possibility of generating an effective magnetic current is provided by adding the second layer, and hence, make the structure three-dimensional (3D). Such a volumetric structure is able to create an array of co-linear electric and magnetic dipole moments in the metasurface plane which is the necessary condition for generating chiral properties.

\begin{figure*}
\centering
   \subfigure[]{
  \epsfig{file=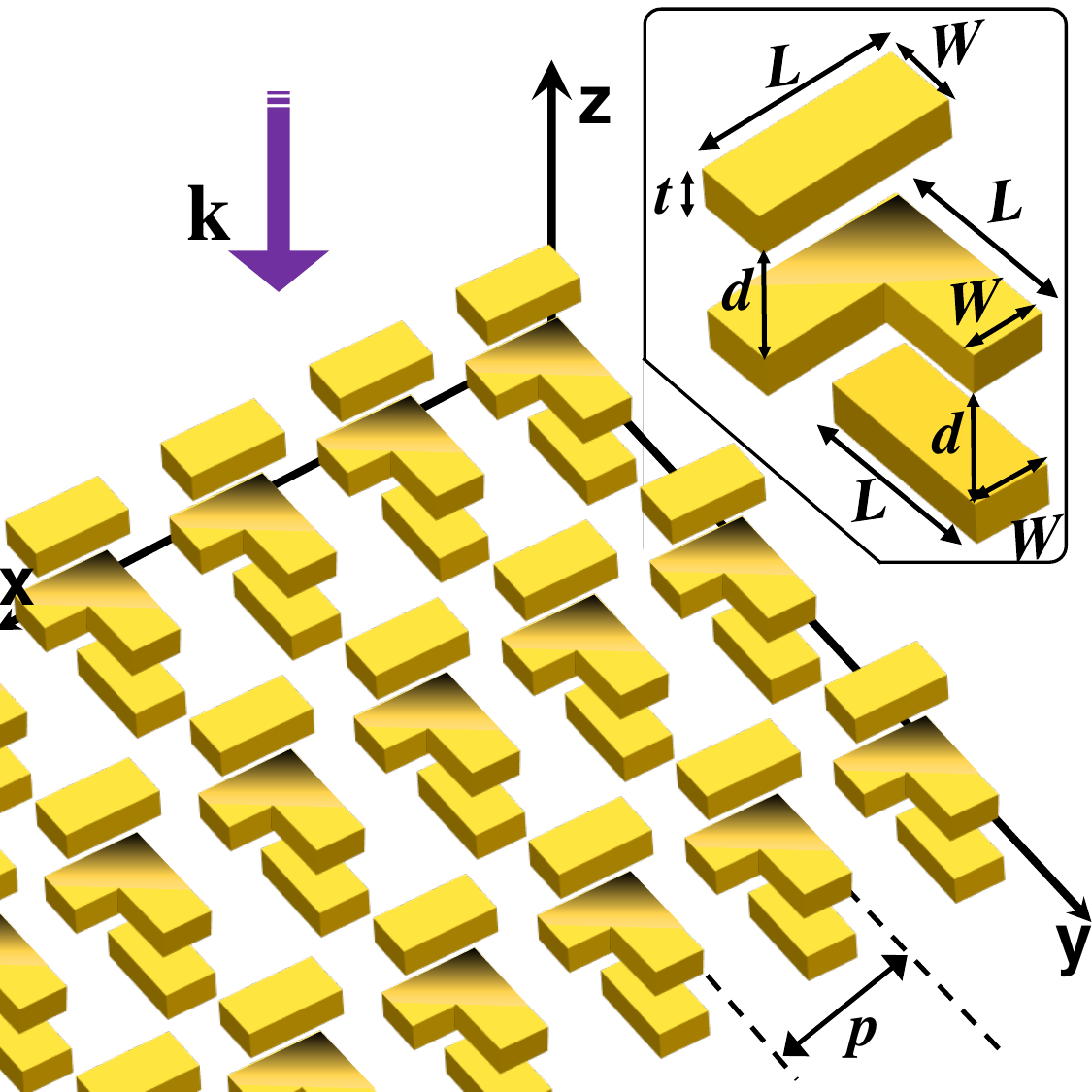, width=0.45\columnwidth}
   \label{ris:figMenzelUa} }
  \subfigure[]{
   \epsfig{file=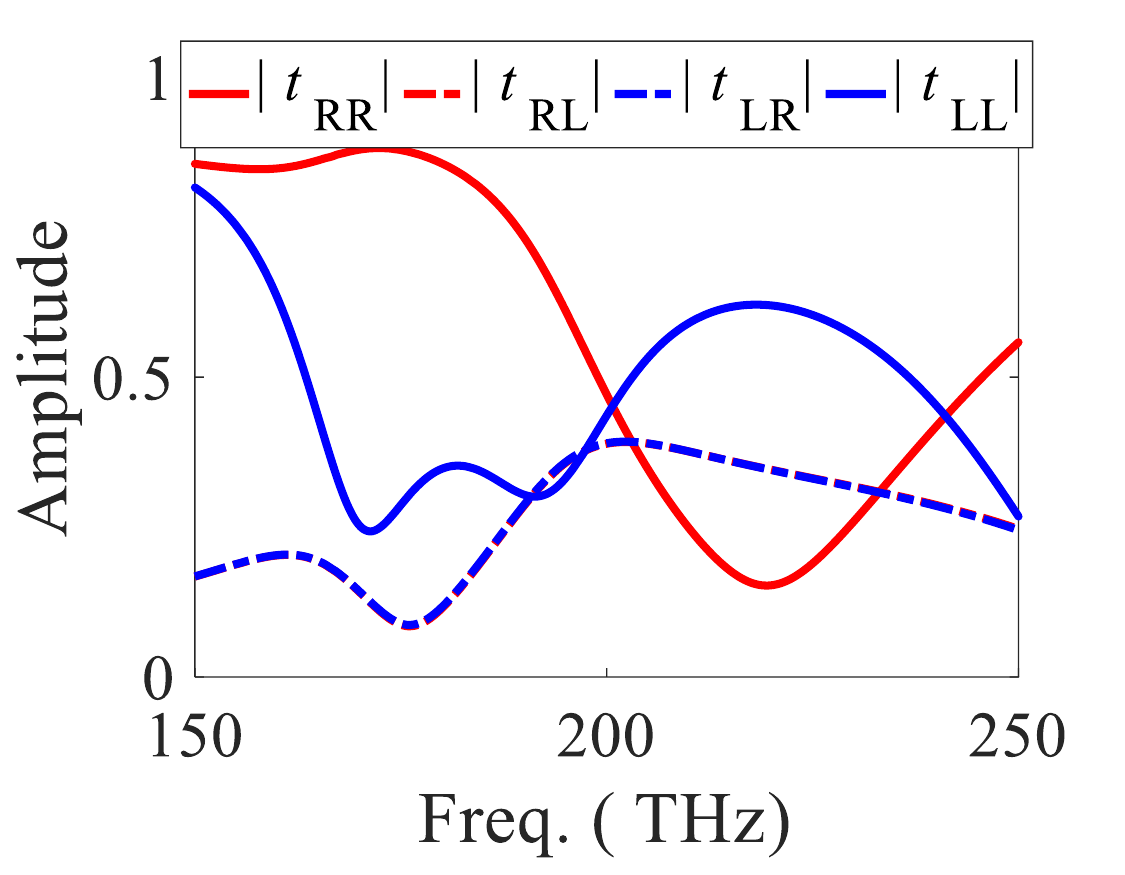, width=0.55\columnwidth}
   \label{ris:figMenzelUb} }
     \subfigure[]{
   \epsfig{file=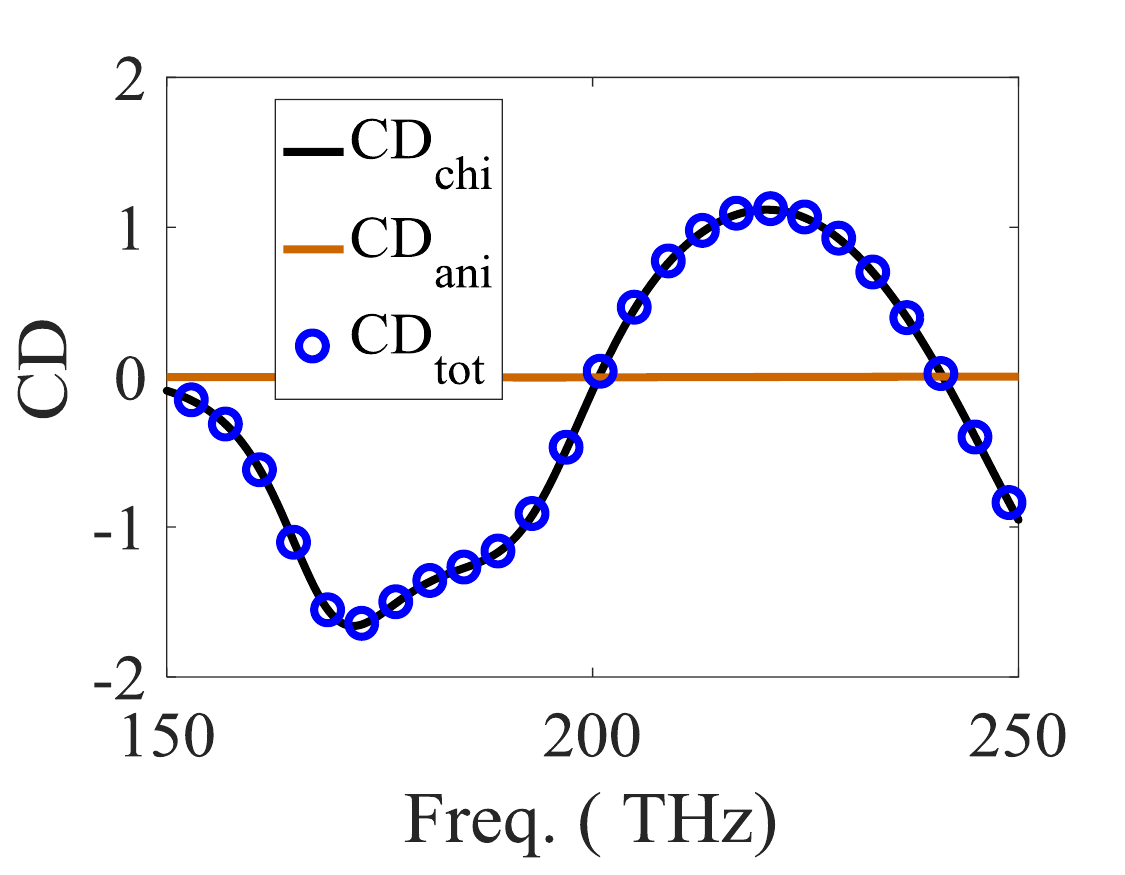, width=0.55\columnwidth}
   \label{ris:figMenzelUc} }
  \caption{(a) A metasurface composed of an array of anisotropic chiral inculsions. The metasurface is composed of three layers: strips in the top and bottom layers and L-shaped particles in the middle. The gold particles are embedded in a the fused silica. Other parameters are $L=290$nm, $W=130$nm, $t=40$nm, $d=80$nm, and $p=500$nm. (b) The amplitude of all components of the transmission coefficients of the proposed metasurface in the CP basis. (c) The contributions of $\rm CD_{ chi}$ and ${\rm CD_{ani}}$ in the observed transmission CD.}\label{ris:figMenzelU}
\end{figure*}

\subsection{A metasurface of L-shpaed plasmonic particles loaded with plasmonic strips on both sides}\label{chirality2D_Rev1:L-shaped2}
So far we have presented two metasurface examples that provide CD of different origins; in the first case CD was originated only from anisotropy properties whereas in the second case CD was originated from both chiral and anisotropy properties of the metasurface. However, as we show here, it is possible to separate the contribution of anisotropy (represented by ${\rm CD_{ani}}$) from the chiral one in the observed CD. Nevertheless, the proposed metasurface yet possesses anisotropy properties which are hidden in the first bracket of the right hand side of Eq.~(\ref{chirality2D_Rev1:CDt_co}), i.e., the ${\eta \left({\alpha}^{\textrm{ee}}_{xx}+{\alpha}^{\textrm{ee}}_{yy}\right)+\left({\alpha}^{\textrm{mm}}_{xx}+{\alpha}^{\textrm{mm}}_{yy}\right)/\eta}$ term when $ {\alpha}^{\textrm{ee}}_{xx}\neq {\alpha}^{\textrm{ee}}_{yy}$~and/or~${\alpha}^{\textrm{mm}}_{xx}\neq {\alpha}^{\textrm{mm}}_{yy}$. Such anisotropy is not recognizable in a CP basis as discussed in Sec.~\ref{CD_exp} and one needs to perform the analysis in an LP basis to recognize it.
\begin{figure}
\centering
   \subfigure[]{
  \epsfig{file=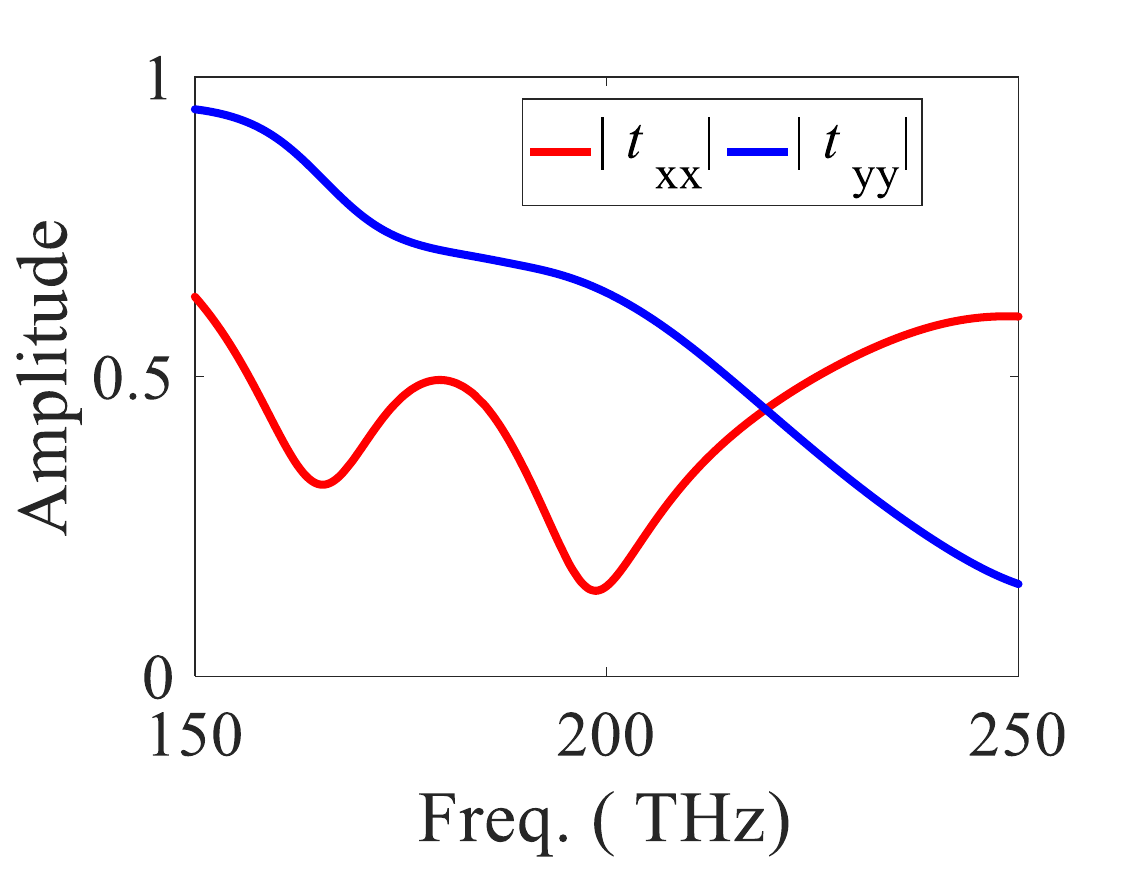, width=0.46\columnwidth}
   \label{ris:figMenzelU_LP2} }
  \subfigure[]{
   \epsfig{file=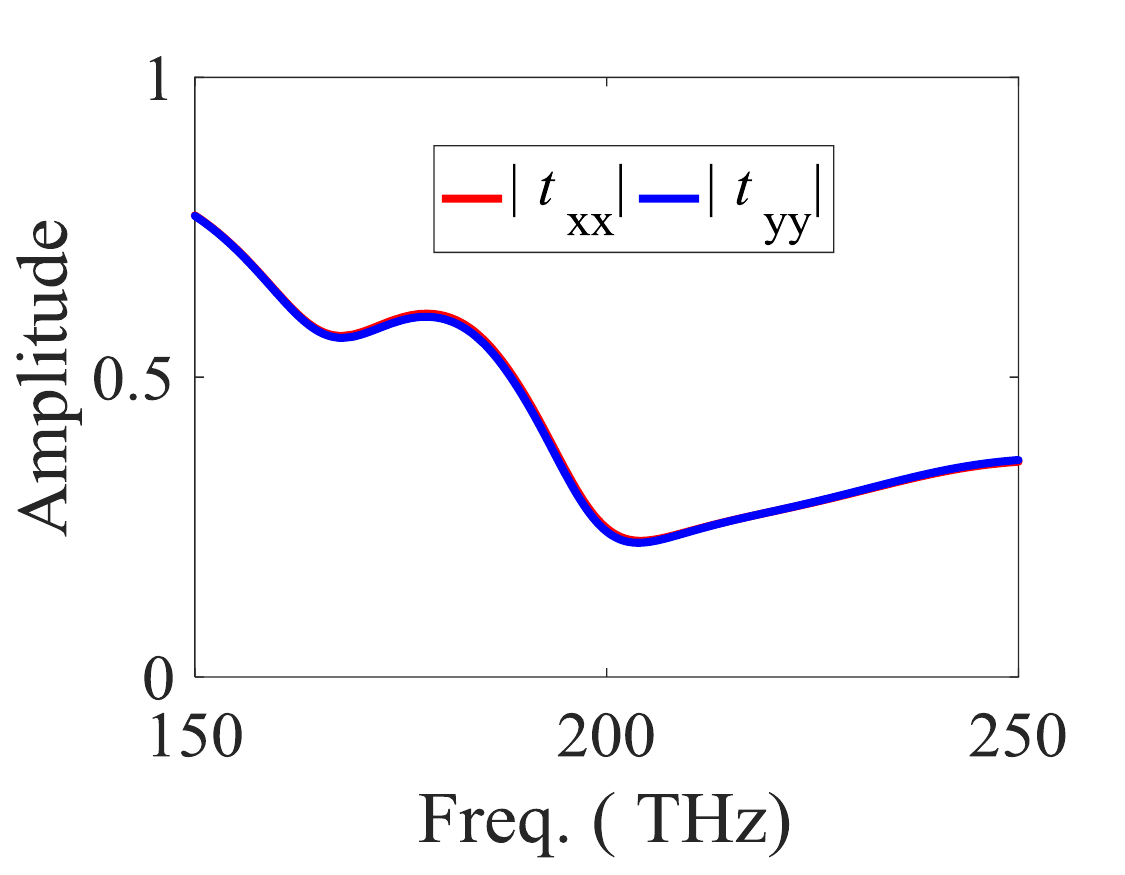, width=0.46\columnwidth}
   \label{ris:figMenzelU_LP1} }
  \caption{The amplitudes of the co-polarized transmission coefficients of the proposed metasurface in Fig.~\ref{ris:figMenzelU} for two different LP bases: (a) when the electric field polarization of the incident basis is along the $x$ and $y$ axes ($\varphi=0$) and (b) when it is rotated by $\varphi=45^\circ$ around the $z$-axis.}\label{ris:figMenzelU_LP}
\end{figure}

To study this illustrative case, evolving from the two previous examples, we consider a metasurface as in Fig.~\ref{ris:figMenzelUa},  composed of a periodic array of resonant scatterers. Each scatterer is composed of an L-shaped gold particle and two strips with equal sizes on the top and bottom of the L-shaped particle. Similarly to what was expressed in Secs.~\ref{anis_weimin1} and~\ref{chi_menzel}, the co- and cross-polarized transmission coefficients in CP basis are plotted in Fig.~\ref{ris:figMenzelUb}. It can be concluded from this figure that the contribution of anisotropy ${\rm CD_{ani}}$ is negligible compared to ${\rm CD_{chi}}$ [see Fig.~\ref{ris:figMenzelUc}]. However, we cannot conclude that the proposed metasurface in this example is isotropic. Indeed, anisotropy of this metasurface is not observed with a CP incident wave or with an LP incident wave with the polarization of its electric field  rotated by $\varphi=45^\circ$ around the $z$-axis [see Fig.~\ref{chirality2D_Rev1:fig1} and Eqs.~(S.25)--(S.28) in Supplementary Material~S--6 for rotation of coordinates]. To clarify this, we have plotted the amplitudes of the co-polarized transmission coefficients for such a metasurface for two different LP incident bases, i.e., when the electric field polarization of the incident basis is along the $x$ or $y$ axis ($\varphi=0$ or $\varphi=90^\circ$), and when it is rotated by $\varphi = 45^\circ$ around the $z$-axis [see Fig.~\ref{ris:figMenzelU_LP}]. As clear from this figure, due to the symmetry of the structure, the  co-polarized transmission coefficients are equal when the incident field is polarized along the two primed directions rotated $\varphi=45^\circ$ around the $z$-axis [see Fig.~\ref{ris:fig1b} for reference to primed directions], whereas they show unequal values for $\varphi=0$. Importantly, this example obviously demonstrates an anisotropy property of the proposed metasurface which was not possible to be recognized from an analysis in CP basis. A simpler example of the issue we are talking about is, e.g., a metasurface composed of a periodic array of resonant dipole scatterers, all oriented along one direction in the metasurface plane (say, e.g., $x$). Such a metasurface is obviously anisotropic and anisotropic properties cannot be recognized in a CP basis since this metasurface responds similarly for both left and right hand CP waves.

\begin{figure*}
\centering
   \subfigure[]{
  \epsfig{file=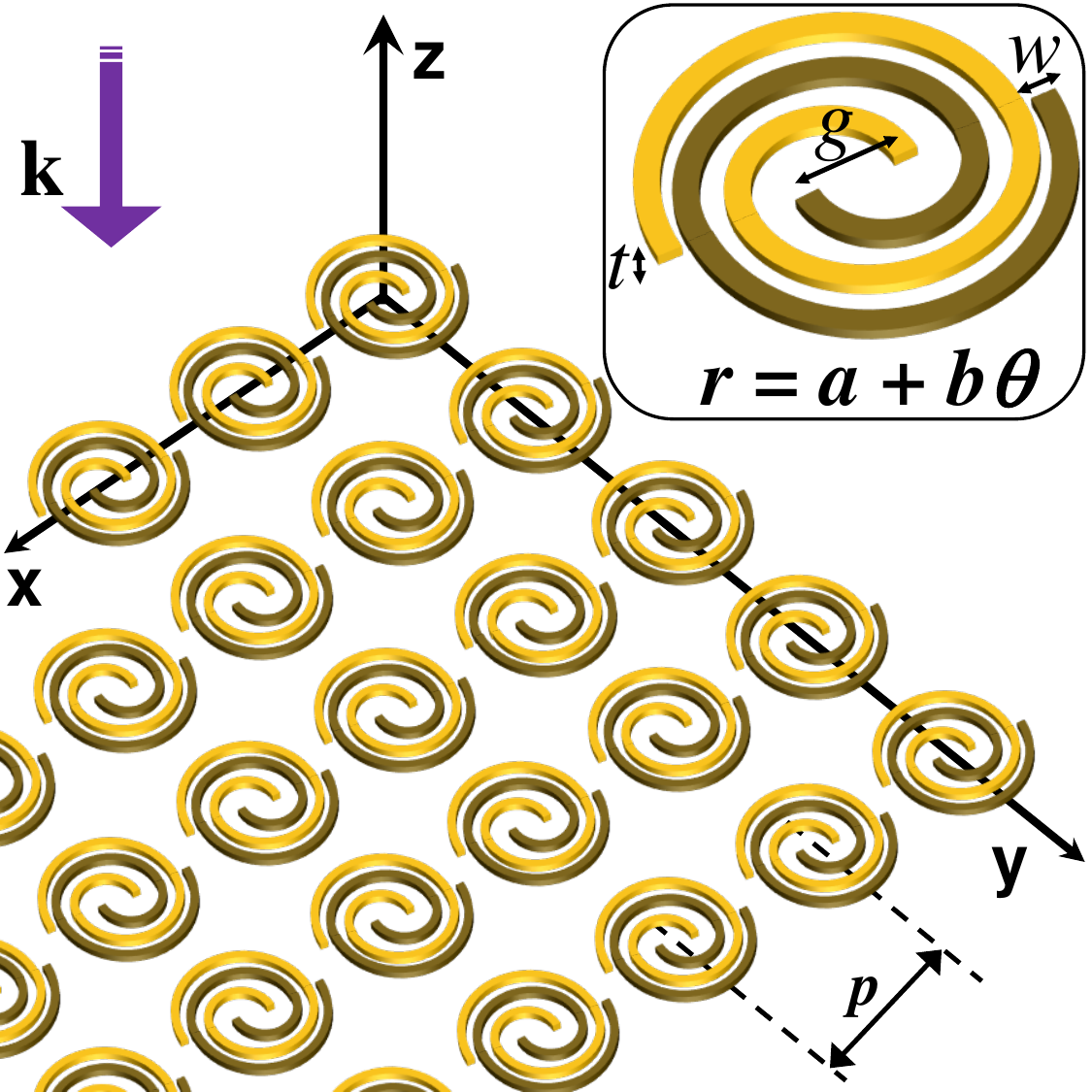, width=0.45\columnwidth}
   \label{ris:figJinweia} }
  \subfigure[]{
   \epsfig{file=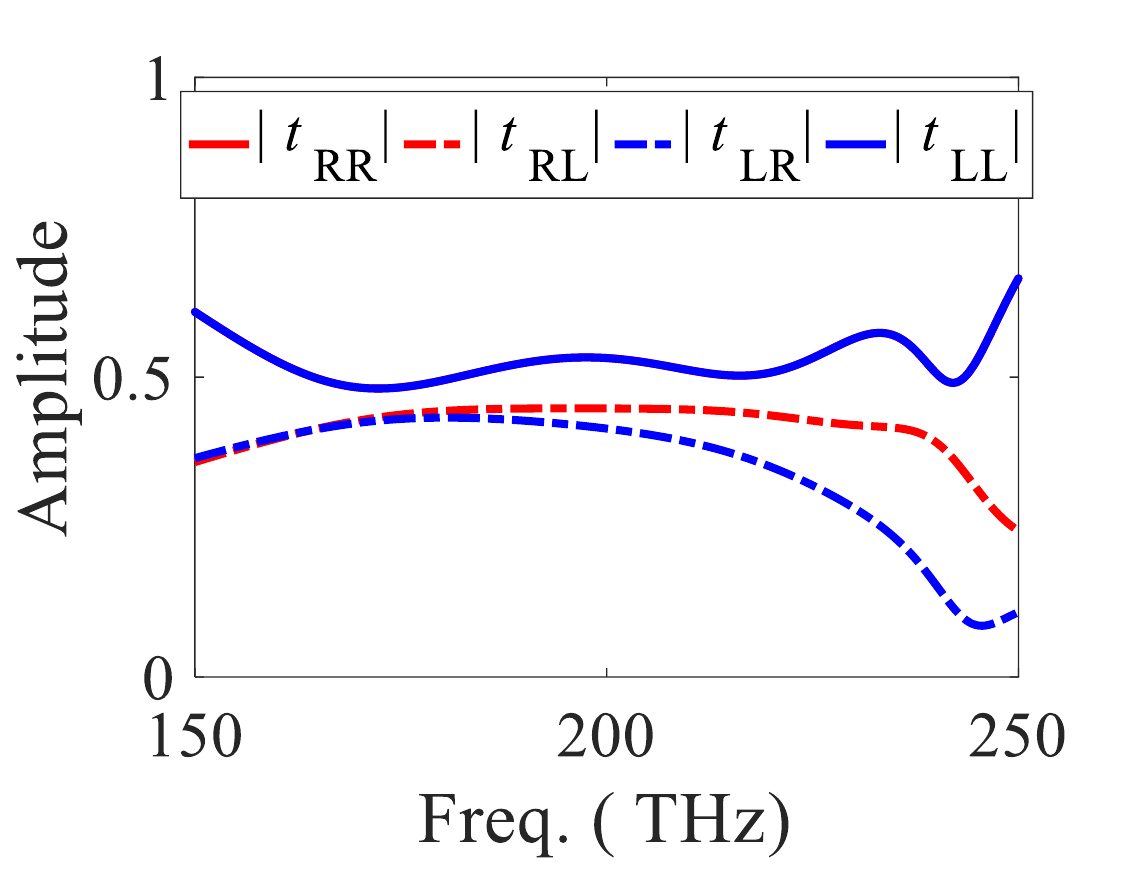, width=0.55\columnwidth}
   \label{ris:figJinweib} }
     \subfigure[]{
   \epsfig{file=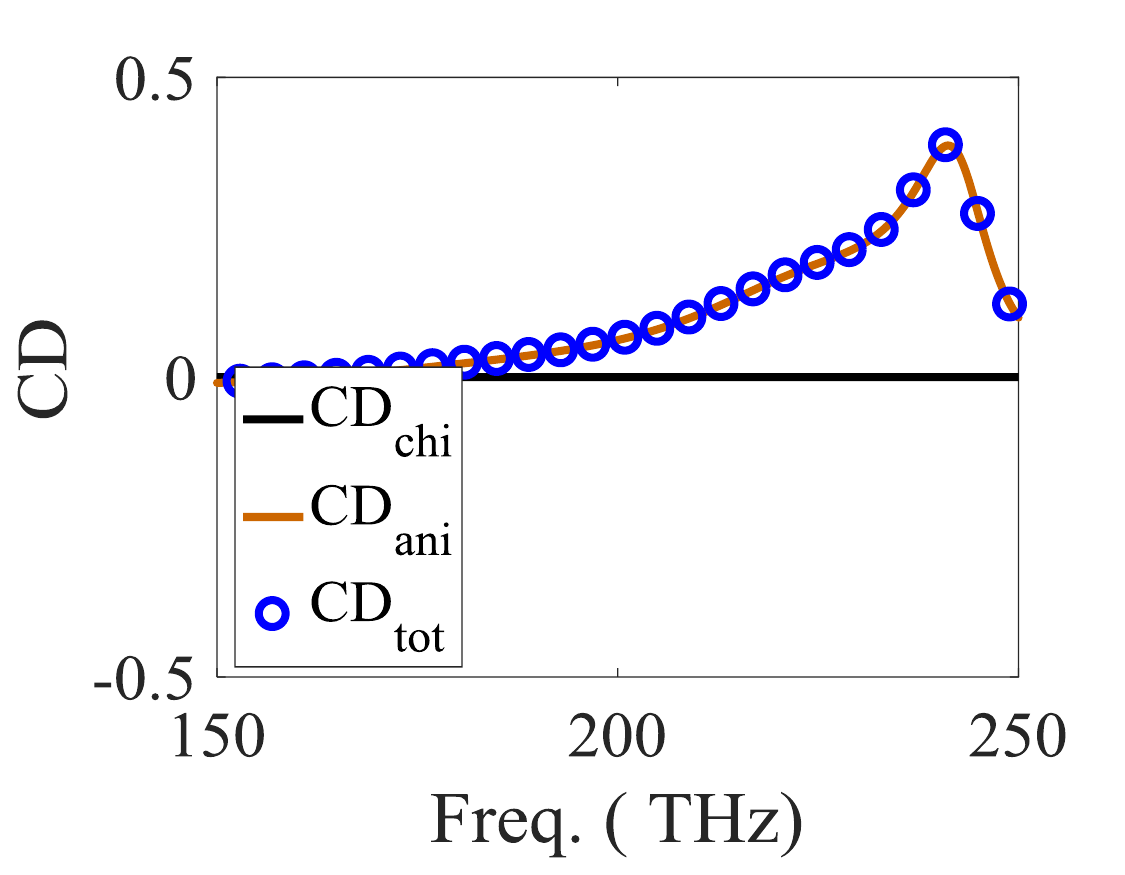, width=0.55\columnwidth}
   \label{ris:figJinweic} }
  \caption{(a) The proposed metasurface composed of infinite number of Archimedean spiral scatterers (see Ref.~\cite{schaferling}) arranged peridocally with period $p=1000~{\rm nm}$. The spiral follows the formula $r=a+b\theta$, where we consider $a=-80~{\rm nm}$ and $b={160\over \pi}~{\rm nm}$. The thickness $t$ and the width $w$ of each spiral are $40$~nm and $80$~nm, respectively, whereas the gap is ${\rm {g}}=165$~nm. (b) The amplitude of transmission coefficients of the proposed metasurface composed of anisotropic double-spiral particles in the CP basis. (c) The contributions of chirality (represented by ${\rm CD_{chi}}$) and anisotropy (represented by ${\rm CD_{ani}}$) in the transmission CD.}\label{ris:figJinwei}
\end{figure*}

\subsection{A metasurface of spiral particles as an achiral anisotropic metasurface with optical activity}\label{metaspi}
In this last example we present a controversial example in the literature, i.e., a metasurface composed of an array of Archimedean spiral scatterers [see Fig.~\ref{ris:figJinweia}]. As suggested in Ref.~\cite{fedotov2006}, such a metasurface should provide chirality, however, as discussed in Sec.~\ref{conorg} no 2D object is intrinsically chiral. Here, we  show that a metasurface composed of such flat inclusions is achiral under normal illumination. The co- and cross-components of the  transmission spectra of such a metasurface are demonstrated in Fig.~\ref{ris:figJinweib}. As it is clear, the EM response of such a metasurface is quite similar to what was discussed in the first example (see Sec.~\ref{anis_weimin1}), i.e., for an anisotropic metasurface. To further prove our statement, the transmission CD of the proposed metasurface together with its CD due to chirality  contribution ${\rm CD_{chi}}$ and anisotropy contribution ${\rm CD_{ani}}$ are sketched in Fig.~\ref{ris:figJinweic}. This clearly demonstrates that the contribution of the metasurface's chirality in the transmission CD is negligible compared to that of the metasurface's anisotropy. Therefore, such a structure should not be considered as a {\it planar} chiral metasurface although it provides a moderate CD values in its transmission spectrum.

As a final remark, it is important to note as discussed in Sec.~\ref{conorg}, although it is  impossible to create a planar chiral structure (with the exception of pseudochirality, observed for oblique incidence), the substrate may induce bianisotropy of any kind such as chiral~\cite{kuwata} or omega~\cite{albooyehSIB1,albooyehSIB2} properties. However, here we have not studied the effect of substrate as an effective way to create chirality since such effect is negligible in case of low contrast refractive indices between the substrate and the superstrate~\cite{albooyehSIB1,albooyehSIB2, albooyehSIB3,albooyehSIB4}. More importantly, the presence of a substrate intrinsically brings about the role of the third dimension and the definition of a {\it planar} structure used in this paper is not valid anymore.

\section{Conclusion}

Based on fundamental principles and analytical modeling, we classified reciprocal metasurfaces and determined chiral, anisotropic, bianisotrpic, isotropic, omega properties. To distinguish among different classes, we observed only the magnitude of reflection and transmission coefficients in linear and/or circular polarization bases. We   presented several expressions to identify each class from reflection and/or transmission amplitude measurements assuming a one-side illumination, which is a typical constraint of   experimental setups.
We provided a few illustrative examples and  elaborate on   possible confusions between two different classes, i.e., anisotropic versus chiral metasurfaces, that seem to occur even in the recent literature.
We have proved that the co-polarized  (cross-polarized) transmission amplitude data can be used to demonstrate the anisotropy of a metasurface in an LP (CP) basis, and they can be used to determine the metasurface chirality in a CP (LP) basis. Furthermore, we emphasize that although anisotropic and chiral metasurfaces can create similar transmission spectra in an LP basis, they can be distinguished based on the fact that the transmission is asymmetric for a chiral metasurface when illuminated from opposite sides whereas it is symmetric for an anisotropic metasurface. The results shown in this paper can be useful when performing CD experiments for metasurfaces.

\begin{acknowledgments}
This work was supported by the W. M. Keck Foundation (USA).
\end{acknowledgments}


\bibliography{main_text}

\end{document}